\documentclass[twocolumn]{aastex61}
\pdfoutput=1 %for arXiv submission
\usepackage{amsmath,amstext}
\usepackage[T1]{fontenc}
\usepackage{apjfonts} 
\usepackage[figure,figure*]{hypcap}

\definecolor{dred}{rgb}{0.75,0,0}
\definecolor{darkred}{rgb}{0.5,0,0}
\definecolor{darkgreen}{rgb}{0,0.5,0}
\definecolor{darkblue}{rgb}{0,0,0.5}
\hypersetup{ colorlinks,
linkcolor=darkblue,
filecolor=darkgreen,
urlcolor=darkred,
citecolor=darkblue}

\newcommand{\beq}{\begin{equation}}
\newcommand{\eeq}{\end{equation}}

\newcommand{\rpp}{r_{\rm p}}
\newcommand{\mzero}{\log M_{0}}
\newcommand{\mone}{\log M_{1}}
\newcommand{\mmin}{\log M_{\rm min}}
\newcommand{\sigmam}{\sigma_{\log \rm M}} 
\newcommand{\wpp}{w_{\rm p}} 

\newcommand{\acen}{\mathcal{A}_{\rm cen}}
\newcommand{\asat}{\mathcal{A}_{\rm sat}}

\shorttitle{AASTeX 6.1 Template}
\shortauthors{Vakili \& Hahn}

\begin{document}

\title{How are galaxies assigned to halos? \\ Searching for assembly bias in the SDSS galaxy clustering}

\author{Mohammadjavad Vakili}
\affiliation{Center for Cosmology and Particle Physics, Department of Physics, New York University, 4 Washington Pl, New York, NY, 10003}
\affiliation{Leiden Observatory, Leiden University, P.O. Box 9513, NL-2300 RA, Leiden, The Netherlands}

\author{ChangHoon Hahn}
\affiliation{Center for Cosmology and Particle Physics, Department of Physics, New York University, 4 Washington Pl, New York, NY, 10003}
\affiliation{Lawrence Berkeley National Laboratory, 1 Cyclotron Rd, Berkeley CA 94720, USA}
\affiliation{Berkeley Center for Cosmological Physics, Campbell Hall 341, University of California, Berkeley CA 94720}
\begin{abstract}
Clustering of dark matter halos has been shown to depend on halo properties beyond mass such as halo concentration, a phenomenon referred to as halo assembly bias. 
%%% 
Standard halo occupation models (HOD) in large scale structure studies assume that halo mass alone is sufficient in characterizing the connection between galaxies and halos. 
%%% 
%%% CHH: @MJV, this sentence is far too long and indirect (RE RefReport Minor #7)
Modeling of galaxy clustering can face systematic effects if the
number of galaxies within a halo is correlated with other halo properties. Using the Small MultiDark-Planck high resolution $N$-body simulation and the clustering measurements of the Sloan Digital Sky Survey (SDSS) DR7 main galaxy sample, we investigate the extent to which the concentration-dependence of halo occupation can be constrained. Furthermore, we study how allowing for the concentration dependence can improve our modeling of galaxy clustering.
%%% 

%%% 
Our constraints on HOD with assembly bias 
suggest that satellite population is not correlated with halo concentration at fixed halo mass. 
%%% 
%%% CHH: @MJV, this sentence is also long and a bit difficult to comprehend -- clarify. 
%%% MJV: FIXED!
At fixed halo mass, our constraints favor lack of correlation between the occupation of centrals and halo concentration in the most luminous samples ($M_{\rm r}<-21.5,-21$), and modest correlation in the $M_{\rm r}<-20.5,-20, -19.5$ samples.
%Furthermore, in terms of the occupation of centrals at fixed halo mass, our results favor lack of correlation with halo concentration in the most luminous samples ($M_{\rm r}<-21.5,-21$), modest levels of correlation for $M_{\rm r}<-20.5,-20, -19.5$ samples.} 
We show that in comparison with abundance-matching mock catalogs, our findings suggest qualitatively similar but modest levels of the impact of halo assembly bias on galaxy clustering. 
The effect is only present in the central occupation and becomes less significant in brighter galaxy samples.
Furthermore, by performing model comparison based on information criteria, we find that in most cases, the standard mass-only HOD model is still favored by the observations.

\end{abstract}


\section{Introduction}
Most theories of cosmology and large-scale structure formation under consideration today rely on the central assumption that galaxies reside in dark matter halos. 
%%%
Detailed study of the galaxy--halo connection is therefore critical in constraining cosmological models (by modeling galaxy clustering at non-linear scales) as well as providing a window into galaxy formation physics. 
%%%
One of the most powerful methods for describing the galaxy--halo connection is the halo occupation 
distribution (HOD, see \citealt{seljak2000, scoccimarro2001,  berlind_weinberg2002, zheng2005, zheng07, leauthaud12, tinker2013, decorated, 2016arXiv160701782H}).

HOD is an empirical framework that provides an analytic prescription for the expected number of galaxies $N$ that reside in halos by specifying a probability distribution function $P(N|x)$ where $x$ is a property of the halo. 
%%%
The standard HOD model assumes that halo mass $M$ alone is sufficient in determining the galaxy population of a halo. 
%%%
%%% CHH: @MJV, what do you mean by "statistical properties of galaxies"? Their spacial statistics? 
%%% MJV: FIXED
In the standard model, the probability of finding $N$ galaxies in a dark matter halo is governed by the halo mass.
%%%
Mathematically, this assumption can be written as $P(N|M,\{x\})=P(N|M)$ where $\{x\}$ is the set of all possible halo properties beyond halo mass $M$.

Despite this simplifying assumption, the models of galaxy--halo connection based on HOD have been successfully used in fitting the measurements of a wide range of statistics such as the projected two-point correlation function of galaxies, small scale redshift space distortion, three-point function, and galaxy--galaxy lensing with remarkable success (e.g. \citealt{zheng07,tinker_rsd2007,zehavi2011,leauthaud12,parejko2013,coupon2015,hod-3pcf,guo2015,miyatake15,zu2015,hod_vs_sham}). 
HOD has been used in constraining the cosmological parameters through modeling the galaxy two-point correlation function (hereafter 2PCF) (\citealt{abazajian2005}), combination of 2PCF with mass-to-light ratio of galaxies (\citealt{tinker05}), redshift space distortions (\citealt{tinker_rsd2007}), mass-to-number ratio of galaxy clusters (\citealt{tinker2012}) galaxy-galaxy weak lensing (\citealt{vdb03,cacciato13,more13,vdb13}) in the main sample of galaxies
of the Sloan Digital Sky Survey (hereafter SDSS, \citealt{york2000}), and also the combination of galaxy clustering and galaxy-galaxy lensing (\citealt{more15}) in SDSS III Baryon Oscillation Spectroscopic Survey (BOSS, \citealt{boss}). Furthermore, HOD is implemented in producing mock galaxy catalogs in the BOSS survey (\citealt{manera2013,white2014}). It has also been used in galaxy evolution studies (\citealt{conroy2009,leauthaud12,behroozi13,hudson2015,zu2015,zu2016}).

The complexity of structure formation however, is not sufficiently modeled under the standard HOD framework. Numerous $N$-body simulations that examine the clustering of 
dark matter halos have demonstrated that halo clustering is correlated with the formation 
history of halos. That is, at a fixed halo mass, the halo bias is correlated 
with properties of halos beyond mass, such as the concentration, formation time, or etc. 
This phenomenon is known as halo assembly bias (see \citealt{sheth2004,gao2005, harker2006, weschler2006, gao2007,croton2007,wang2007,angulo2008,dalal2008,li2008,sunayama2016}). It has been claimed that there is support for halo assembly bias in observations of SDSS redMaPPer galaxy clusters (\citealt{miyatake2016}).

Furthermore, the halo occupation may also depend on the formation history of halos. Then we may expect the spatial statistics of galaxies to be tied to the halo properties beyond mass such as the concentration of halos. There have been many attempts in the literature at examining the dependence of halo occupation on environment of the halos. But the results are mixed, and there is very little consensus. \citet{tinker_void_2009} show that the properties of the galaxies that reside in voids can be explained by the halo mass in which they live, and their properties are independent of their large scale environment of the halos. \citet{tinker_density_hod} proposes an extension of the standard HOD model $P(N|M)$ such that the number of galaxies residing in a halo not only depends on the mass of the halo, but also on the large scale density contrast $P(N|M,\delta)$. Based on modeling the clustering and void statistics of the SDSS galaxies, \citet{edHOD-tinker} shows that the dependence of the expected number of central galaxies on large scale density is not very strong.    
By randomly shuffling the galaxies among host halos of similar mass in the Millenium simulation, \citet{croton2007} shows that assembly bias significantly impacts the galaxy two-point correlation functions. They also show that the effect is different for the faint and the bright samples.

%Based on a group catalog of SDSS galaxies, \citet{blanton_group_2007} find that shuffling of galaxy colors among groups of similar mass has very little effect (only a slight change in small scales: less than $\sim 300 \rm{kpc}$) on the clustering of galaxies. Based on an alternative group catalog, \citet{wang2013} shows that clustering of the central galaxies does not depend only on the halo mass, but also on additional parameters such as the specific star formation rate.
%\todo{I think this paragraph might be a bit misleading to the reader, I think the paper could do without it. %Your work doesn't incorporate galaxy properties, but only halo properties.}

Another family of empirical galaxy--halo connection models is \emph{Abundance} \emph{Matching}. In Abundance Matching models, galaxies are assumed to live in halos and are assigned luminosities or stellar masses by assuming a monotonic mapping. In this monotonic mapping, the abundance of the halos are matched to the abundance of some property of galaxies (\citealt{kravtsov2004,vale2004,tasitsiomi2004,conroy2009,guo2010,wetzel2010,Neisten2011,watson2012,rodriguez2012,kravstov2013,mao2015,chavez2016}). 
One of the most commonly used host halo properties in abundance matching is the maximum circular velocity of the host halo $V_\mathrm{max}$ that traces the depth of the gravitational potential well of the halo. Furthermore, a scatter is assumed in this mapping. Within this galaxy--halo connection framework, abundance matching models have been successfully used in modeling a wide range of the statistical properties of galaxies such as two-point correlation function (\citealt{reddick2013,lehman2015,hod_vs_sham}) as well as the group statistics of galaxies (\citealt{sham_gmf}). 

It has been shown that the abundance matching mock catalogs that use $V_{\rm max}$ (see \citealt{hw2013,arz2014}), or the ones that use some combination of $V_{\rm max}$ and host halo virial mass $M_{\rm vir}$ (see \citealt{lehman2015}) exhibit significant levels of assembly bias. 
That is, halo occupation in these models depends not only on halo mass, but also on other halo properties. This has been demonstrated by randomizing the galaxies among host halos in bins of halo mass, such that the HOD remains constant, and then comparing the difference in the 2PCF of the randomized catalog and that of the original mock catalog. 

Based on the projected 2PCF measurements of (\citealt{hw2013}) galaxy catalogs, \citet{arz2014} showed that after fitting the 2PCF measurements of these catalogs with the standard \emph{mass-only} HOD modeling, the inferred HOD does not match the \emph{true} halo occupation of these catalogs. That is, in the presence of assembly bias in a galaxy sample, one can fit the clustering of this sample with the standard \emph{mass-only} HOD, but that does not guarantee recovery of the true HOD parameters.

%In this work, we aim to investigate the dependence of halo occupation on halo concentration, and how this dependence can be constrained in the low-redshift universe with the measurements of the 2PCF of galaxies with a wide range of luminosities in the SDSS DR7 main galaxy sample. 
%%% CHH: Re-phrasing 
In this work, we aim to investigate the dependence of halo occupation on 
    halo concentration and how this dependence can be constrained in the low-redshift universe by 2PCF measurements of galaxies in a wide range of luminosities in the SDSS DR7 main galaxy sample. 
%%%
In order to achieve this goal, we need to adopt a HOD model that takes into account 
a dependence on halo properties beyond mass. 
%%% CHH: Re-phrasing to address ref. report
    A number of frameworks in the literature \citep{edHOD-tinker,edHOD-gillmartin,edHOD-weinberg} 
    have proposed environment-dependent HOD models that take into account the large-scale density contrast. 
    In this investigation, we use the following case of the decorated HOD framework \citep{decorated}.  
    In our decorated HOD framework, at fixed halo mass, halos are populated with galaxies according to 
    the standard HOD model. Then using a secondary halo property, halos are split into two populations
    in halo mass bins: halos with the highest and lowest secondary property values. Afterwards, based
    on the assembly bias amplitude parameter, the number of galaxies in the two populations are enhanced
    or reduced. In this model, the assembly bias parameter is not degenerate with the rest of the HOD 
    parameters. 
%%%

The advantage of this framework is that the more complex HOD model is identical to the underlying \emph{mass-only} HOD model in every respect, except that at a fixed halo mass, halos receive enhancement (decrements) in the number of galaxies they host according to the value of their secondary property. In order to constrain assembly bias along with the rest of the HOD parameters, we make use of the publicly available measurements of the projected 2PCF and number density measurements made by (\citealt{guo2015}). These measurements made use of the NYU Value-Added Galaxy Catalog (\citealt{Blanton2005}).

Furthermore, we discuss how taking assembly bias into account in a more complex HOD model can improve our modeling of galaxy clustering in certain brightness limits. Then, we make a qualitative comparison between the levels of the impact of assembly bias in our best-fit decorated HOD model on galaxy clustering, and the impact of assembly bias present in (\citealt{hw2013}) catalogs on galaxy clustering. Our comparison shows the levels of the impact of assembly bias on galaxy clustering seen in the predictions of both models follow the same trend. That is assembly bias is more prominent in lower luminosity-threshold samples and its impact on galaxy clustering is only significant on large scales (more than a few Mpc).

In order to investigate whether the additional complexity of the decorated HOD model is demanded by the galaxy clustering data, we perform a model comparison between the standard HOD model and the HOD model with assembly bias. We also discuss the effect of our choice of $N$-body simulation on our constraints, and previous works in the literature (\citealt{zentner2016}) based on smaller $N$-body simulations. In addition to analysis of the luminosity-threshold samples presented in \citet{zentner2016}, we consider the brightest ($M_{\rm r}<-21.5$) galaxy sample. For the samples considered in both \citet{zentner2016} and this investigation, we compare the constraints on the expected levels of assembly bias.

This paper is structured as follows: In Section \ref{sec:method} we discuss the $N$-body simulation, the two halo occupation modeling methods, and the details of the computation of model observables used in this investigation. Then in Section \ref{sec:data}, we discuss the data used in this study. 
In Section \ref{sec:analysis} we discuss the details of our inference analysis as well as the results. This includes description of the details of our inference setup. In Section \ref{sec:result} we discuss the constraints and their implications. This includes presentation of the constraints on the parameters of the two models, interpretation of the predictions of our constraints and their possible physical ramifications, assessment of the levels of assembly bias as predicted by our model constraints and its comparison with abundance matching mock catalogs, and finally model comparison. Finally, we discuss and conclude in Section \ref{sec:summary}.
Throughout this paper, unless stated otherwise, 
all radii and densities are stated in comoving units. 
Standard flat $\Lambda$CDM is assumed, and all cosmological 
parameters are set to the Planck 2015 best-fit estimates.

\section{Method}\label{sec:method}

In this section, we discuss the ingredients of our modeling one-by-one. First, we discuss the simulation used in this study. Afterwards, we talk about the forward modeling of galaxy catalogs in the standard HOD modeling framework as well as the decorated HOD framework. Then, we provide an overview of the two summary statistics of the galaxy catalogs that we used in our inference.  

\subsection{Simulation}

For the simulations used in this work, we make use of the Rockstar (\citealt{rockstar}) halo catalogs in the $z=0$ snapshot of the Small MultiDark of Planck cosmology (hereafter $\mathtt{SMDP}$; \citealt{smallmultidark}). 
This high resolution $N$-body simulation\footnote{publicly available at \url{https://www.cosmosim.org}}
was carried out using the 
%%% CHH: small typo
GADGET-2 code (see \citealt{smallmultidark} 
and references therein), 
%%%
following the Planck $\Lambda$CDM cosmological parameters 
$\Omega_{\rm m} = 0.307$, $\Omega_{\rm b} = 0.048$, $\Omega_{\Lambda} = 0.693$, $\sigma_{8} = 0.823$, $n_{\rm s}=0.96$, 
$h=0.678$. It has a box size of 
$(0.4\,h^{-1} \rm{Gpc})^3$  with 3840$^3$ simulation particles with 
$m_{\rm p} = 9.6 \times 10^{7} \; h^{-1} M_{\odot}$ and 
gravitational softening length of $\epsilon = 1.5\,h^{-1} \rm{kpc}$.

%%% CHH: reworded for clarity
    The Rockstar algorithm, in SMDP, can reliably 
    resolve halos with $\geq 100$ particles, which corresponds to 
    $M_{\rm vir} \geq 9.6 \times 10^{9} \;  h^{-1}M_{\odot}$~\citep{halodemographic}.
%In the $\mathtt{SMDP}$ simulation, as discussed in \citet{halodemographic}, the Rockstar algorithm can reliably resolve halos with $\geq 100$
%%% CHH: It was a bit wordy before 
%%{\bf \color{purple} particles, which corresponds to $M_{\rm vir} \geq 9.6 \times 10^{9} \; h^{-1}M_{\odot}$.} 
    The $\mathtt{SMDP}$ simulation satisfies both the size and 
    resolution requirements of studying the galaxy--halo connection 
    in a wide range of luminosity thresholds.
%The $\mathtt{SMDP}$ simulation provides a number of advantages by satisfying both the size  and resolution requirements of studying the galaxy--halo connection in a wide range of  luminosity thresholds.
     For fainter galaxy samples, the faintest galaxies reside in lower 
    mass halos, which requires high resolution. Meanwhile for luminous galaxy samples, their 
    lower number densities requires a large comoving volume. 

%%%
%The advantage of using the $\mathtt{SMDP}$ simulation is that it satisfies both the size and resolution 
%requirements of studying the galaxy--halo connection in a wide range of luminosity thresholds. 
%%%

%%%
Furthermore, since we are studying the higher order halo occupation statistics, the concentration-dependence in particular, it is important to use a simulation that can resolve the internal structure of halos. 
% CHH: I think it sounds more fluid here rather than at the end of the section. I've also reworded things 
    We only keep the Rockstar halos with more than 
    $1000$ particles, for which the halo concentration can be determined 
    reliably. This corresponds to a minimum viral mass of 
    $M_{\rm vir} \geq 9.6 \times 10^{10} \; h^{-1}M_{\odot}$. 
    Based on this halo mass cut, we exclude the faintest DR7 
    galaxies samples from our analysis.
%{\bf \color{purple} Since we are interested in constraining the concentration-dependent decorated HOD model, we only keep the Rockstar halos with more than $1000$ particles, for which the halo concentration can be determined reliably. This corresponds to a minimum viral mass of $M_{\rm vir} \geq 9.6 \times 10^{10} \; h^{-1}M_{\odot}$. As a result of this minimum cut on halo mass, we will exclude the faintest DR7 galaxies samples from our analysis.} 

%%% CHH: Re-worded based on (github issue #1)
    In the context of Subhalo-Abundance Matching models, which requires subhalo completeness 
    in the low mass limit, the $\mathtt{SMDP}$ simulation has been used to model the faintest 
    galaxy samples in the SDSS data~\citep[see][]{hod_vs_sham}. 

%%%
The added advantage of using the $\mathtt{SMDP}$ simulation over some of the other industrial simulation boxes commonly used in the literature such as $\mathtt{Bolshoi}$ (\citealt{Klypin2011,smallmultidark}) simulation is its larger comoving volume. Larger volume makes this simulation more suitable for performing inference with $L_{\star}$ (corresponding to $M_{\rm r} \sim -20.44$, see \citealt{blanton2003}) and more luminous than $L_{\star}$ galaxy samples that occupy larger comoving volumes.

\subsection{Halo occupation modeling}
\subsubsection{standard model without assembly bias}\label{subsubsec:hod}

For our standard HOD model, we assume the HOD parameterization from \citet{zheng07}. 
According to this model, a dark matter halo can host a central galaxy and some number of satellite galaxies. The occupation of the central galaxies follows a nearest-integer distribution, 
and the occupation of the satellite galaxies follows a Poisson distribution. The expected number of centrals and satellites as a function of the host halo mass of $M_{h}$ are given by the following equations 
\begin{eqnarray}
\langle N_{\rm c}| M_{h} \rangle &=& \frac{1}{2}\Big[1+\Big(\frac{\log M_{h} - \log M_{\rm min}}{\sigma_{\log \rm{M}}} \Big) \Big], \label{hod:central}\\ 
\langle N_{\rm s} | M_{h} \rangle &=& \Big( \frac{M_{h} - M_{\rm{0}}}{M_{\rm 1}} \Big)^{\alpha}. \label{hod:satellite}
\end{eqnarray}

For populating the halos with galaxies, we follow the procedure described in \citet{2016arXiv160701782H}, and \citet{decorated}. The central galaxies are assumed to be at the center of the host dark matter halos. We assume that the central galaxies are at rest with respect to the bulk motion of the halos and their velocities are given by the velocity of the center of mass of their host halo. Note that this assumption is shown to be violated in brighter than $L_{\star}$ galaxy samples (see \citealt{guo2015}). But since we are not considering the redshift space 2PCF multipoles in our study, we do not expect this velocity bias to impact our inference. We place the satellite galaxies within the virial radius of the halo following a Navarro-Frenk-White profile (hereafter NFW; \citealt{nfw}). This approach is different from other simulation-based halo occupation modeling techniques (see \citealt{hod_vs_sham,zheng_guo}) in that the positions of the satellites are not assigned to the dark matter particles in the $N$-body simulation. 
%%% Missing citation
%%%MJ: FIXED! Concentration values are read directly from the N-body halo catalog.
    The concentration of the NFW profile comes directly from the halo catalog. 
%%%
The velocities of the satellite galaxies are given by two components. The first component is the velocity of the host halo. The second component is the velocity of the satellite galaxy with respect to the host halo which is computed following the solution to the NFW profile Jeans equations (\citealt{more2010}). We refer the readers to \citet{decorated} for a more comprehensive and detailed discussion of the forward modeling of the galaxy mock catalogs.

\subsubsection{Model with Assembly bias}\label{subsubsec:decorated}
Now let us provide a brief overview of HOD modeling with $\mathtt{Heaviside}$ $\mathtt{Assemblybias}$ (referred to as the decorated HOD) introduced in \citet{decorated}. At a fixed halo mass $M_{h}$, halos are split into two populations: population of halos with the 0.5-percentile of highest concentration, and population of halos with 0.5 percentile of lowest concentration. For simplicity, we call the first population ``type-1'' halos, and the second population ``type-2'' halos. In the decorated HOD model, the expected number of central and satellite galaxies at a fixed halo mass $M_{h}$ in the two populations are given by
\begin{eqnarray}
\langle N_{c,i} | M_{h},c\rangle &=& \langle N_{c} | M_{h}\rangle + \Delta N _{c,i}, \; i=1,2 \label{eq:decoratedcentral} \\
\langle N_{s,i} | M_{h},c\rangle &=& \langle N_{s} | M_{h}\rangle + \Delta N _{s,i}, \; i=1,2 \label{eq:decoratedsatellite}
\end{eqnarray}
where $\langle N_{c} | M_{h}\rangle$ and $\langle N_{s} | M_{h}\rangle$ are given by Eqs \ref{hod:central} and \ref{hod:satellite} respectively, and we have $\Delta N_{s,1} + \Delta N_{s,2} = 0$, and $\Delta N_{c,1} + \Delta N_{c,2} = 0$. These two conditions ensure the conservation of HOD. At a given host halo mass $M_{h}$, the central occupation of the the two populations follows a nearest-integer distribution with the first moment given by \ref{eq:decoratedcentral}; and the satellite occupation of the the two populations follows a Poisson distribution with the first moment given by \ref{eq:decoratedsatellite}.

In this occupation model, the allowable ranges that quantities $\Delta N_{c,1}$ and $\Delta N_{s,1}$ can take are given by 

\begin{eqnarray}
\Delta N_{c,i,\rm min} \leq &\Delta N_{c,i}& \leq \Delta N_{c,i,\rm max}
 , \label{eq:cen-bounds} \\
 \Delta N_{c,,\rm min} &=& \mathrm{max} \{-\langle N_{c} | M_{h}\rangle, \langle N_{c} | M_{h}\rangle -1 \}, \\ 
 \Delta N_{c,i,\rm max} &=& \mathrm{min} \{\langle N_{c} | M_{h}\rangle, 1-\langle N_{c} | M_{h}\rangle\}, \\
-\langle N_{s} | M_{h}\rangle \leq & \Delta N_{s,i}& \leq \langle N_{s} | M_{h}\rangle. \label{eq:sat-bounds}
\end{eqnarray}
Afterwards, the assembly bias parameter $\mathcal{A}$ is defined in the following way:

\begin{eqnarray}
\Delta N_{\alpha , 1}(M_{h}) &=& |\mathcal{A_\alpha}| \Delta N_{\alpha , 1}^{\rm max}(M_{h}) \; \; \rm{if} \; \mathcal{A_\alpha} > 0,  \label{positive_A} \\
\Delta N_{\alpha , 1}(M_{h}) &=& |\mathcal{A_\alpha}| \Delta N_{\alpha , 1}^{\rm min}(M_{h}) \; \; \rm{if} \; \mathcal{A_\alpha} < 0, \label{negative_A}
\end{eqnarray}
where the subscript $\alpha = c , s$ stands for the centrals and satellites respectively, and $\Delta N_{\alpha , 1}^{\rm max}(M_{h})$, $\Delta N_{\alpha , 1}^{\rm min}(M_{h})$ are given by Eqs. \ref{eq:cen-bounds} and \ref{eq:sat-bounds}. 

For a given $\mathcal{A}_{\alpha}$, once $\Delta N_{\alpha,1}$ is computed using equation (\ref{positive_A})---if $\mathcal{A}_{\alpha}>0$---or equation (\ref{negative_A})---if $\mathcal{A}_{\alpha}<0$---, $\Delta N_{\alpha,2} = 1 - \Delta N_{\alpha,1}$ is computed. At a fixed halo mass $M_{h}$, once the first moments of occupation statistics for the $type$-1 and $type$-2 halos are determined, we perform the same procedure described in \ref{subsubsec:hod} to populate the halos with mock galaxies.

\subsubsection{Redshift-space distortion}

Once the halo catalogs are populated with galaxies, the real-space positions and velocities of all mock galaxies are obtained. The next step is applying a redshift-space distortion transformation by assuming plane-parallel approximation.
Our use of plane parallel approximation is justified because of the narrow redshift range of the SDSS main galaxy sample considered in this study.
If we assume that the $\hat{z}$ axis is the line-of-sight direction, then with the transformation $(X,Y,Z) \rightarrow (S_x,S_y,S_z) = (X , Y ,Z + v_{z}(1+z)/H(z))$
for each galaxy with the real space coordinates $(X,Y,Z)$, velocities $(v_x,v_y,v_z)$, and redshift $z$, we obtain the redshift-space coordinate of the produced mock galaxies. Here we assume $z \simeq 0$, and therefore transformation is given by $(X,Y,Z) \rightarrow (X,Y,Z+v_{z}/H_{0})$.   

\subsection{Model Observables}

As described in \citet{decorated} and \citet{2016arXiv160701782H}, this approach makes no appeal to the fitting functions used in the analytical calculation of the 2PCF. The accuracy of these fitting functions is limited (\citealt{tinker08,tinker10,watson13}). Our approach also does not face the known issues of the treatment of halo exclusion and scale-dependent bias that can lead to potential inaccuracies in halo occupation modeling (see \citealt{vdb13}). 

The projected 2PCF $w_{p}(r_{p})$ can be computed by integrating the 3D redshift space 2PCF $\xi(r_{p} , \pi)$ along the line-of-sight (where $r_{\rm p}$ and $\pi$ denote the projected and line-of-sight separation of galaxy pairs respectively):
\beq
w_{p}(r_{p}) = 2 \int_{0}^{\pi_{max}}\xi(r_{p} , \pi)\; d\pi
\label{los}
\eeq
For our 2PCF calculations, we use the $w_{\rm p}$ measurement 
functionality of the fast and publicly available pair-counter 
code 
%%% CHH: I think it's betters to include URLs as footnotes
    $\mathtt{CorrFunc}$\footnote{available at \url{https://github.com/manodeep/Corrfunc}}
(\citealt{corrfunc}). To be consistent with the SDSS measurements described in Section \ref{sec:data}, $w_{\rm{p}}(r_{\rm{p}})$ is obtained by the line-of-sight integration to $\pi_{max}=40 \; h^{-1}\rm{Mpc}$. Note that $w_{p}(r_{p})$ is measured in units of $h^{-1}\rm{Mpc}$. To be consistent with \citet{guo2015}, we use the same binning (as specified in Section \ref{sec:data}) to measure $w_{\rm p}$. In addition to the projected 2PCF, we use the number density given by the number of mock galaxies divided by the comoving volume of the $\mathtt{SMDP}$ simulation. 

Note that a full forward model of the data requires running the simulation at different redshifts, generation of light cones, accounting for the complex survey geometry and systematic errors such as fiber collisions. Using the $z=0$ output of the $\mathtt{SMDP}$ simulation in our forward model of the spatial distribution of galaxies is only an approximation. This approximation can be justified by the small redshift range of the SDSS DR7 main galaxy sample. As described in \citealt{zehavi2011}, using random catalogs with angular window function of the data in measurements of galaxy clustering accounts for the geometry of the data. As described in Section~\ref{sec:data}, the fiber collision correction method of \citealt{guo2012} is applied to the SDSS clustering measurements used in this study. Therefore we do not account for that effect in our forward model.

\section{Data}\label{sec:data}

We focus on the measurements made on the volume-limited luminosity-threshold main sample of galaxies in the SDSS spectroscopic survey. In this section, we briefly describe the measurements used in our study for finding constraints on the assembly bias as well as the HOD parameters.

The measurements consist of the number density $n_{g}$ and the projected 
 %%% CHH: minor rewording 
    2PCF, $w_{p}(r_{p})$, from 
\citet{guo2015} for the volume-limited sample of galaxies in NYU Value Added Galaxy Catalog (\citealt{Blanton2005}) constructed from the SDSS DR7 main galaxy sample (\citealt{abazajian2009}). 
%%% CHH: shortened
    In particular, \cite{guo2015} constructed eight volume-limited
    luminosity-threshold samples with maximum $M_r$ of 
    -18, 18.5, -19, -19.5, -20, -20.5, -21, and -21.5. 
Qualitatively, these samples are constructed in a similar way to those constructed in \citet{zehavi2011}. For detailed differences between the samples in \citet{guo2015} and \citet{zehavi2011}, we refer the reader to the Table 1 and Table 2 in those papers respectively. 
%%% CHH: fixed some typos and reworded
    As pointed out in Section~\ref{sec:method}, we apply a lower mass 
    cut of $M_{\rm vir} \sim 9.6 \times 10^{10} \; h^{-1}M_{\odot}$ 
    to the halo catalog. Since galaxies in the faintest 
    luminosity-threshold samples ($M_{\rm r}<$-18, -18.5, -19) can 
    reside in halos below this halo mass cut, we exclude them from 
    the rest of this analysis.

The projected 2PCFs are measured in 12 logarithmic $r_{p}$ bins (in units of $h^{-1}\rm{Mpc}$) of width $\Delta \log(r_{\rm p}) = 0.2$, starting from $r_{\rm{p}} = 0.1 \; h^{-1}\rm{Mpc}$. For all luminosity threshold samples, the integration along the line-of-sight (\ref{los}) are performed to $\pi_{max}=40 \; h^{-1}\rm{Mpc}$. 

The 2PCF measurement of each luminosity-threshold sample is accompanied by a covariance matrix constructed using 400 jackknife sub-samples of the data. The number density measurements are also accompanied by uncertainties measured using the jackknife method. Furthermore, the covariance between the number denisty and the projected 2PCF measurements are neglected. As \citet{norberg} shows, parameter estimation using jackknife covariance matrices is conservative as the jackknife method overestimates the errors in the observations. 

%%% CHH: reworded for clarity
The jackknife covariances from \citet{guo2015} only capture 
uncertainties in the observations. However, our modeling approach, 
which is based on populating an $N$-body simulation with a finite 
volume, also has uncertainties comparable to the uncertainties in 
the observations. 
%Our modeling approach is based  on populating an $N$-body simulation with a finite volume. Therefore, our model has uncertainties that can be comparable to the measurement uncertainties. 
Consequently, the covariances that come from the model uncertainties
$\widehat{C}_{\rm m}$ need to be added to the estimate of the 
covariance matrix of the observations $\widehat{C}_{\rm d}$:
\beq
\widehat{C} = \widehat{C}_{\rm d} + \widehat{C}_{\rm m}. 
\eeq
In order to estimate the model uncertainties, we follow the 
prescription described in \citet{zheng_guo,guo2015,hod_vs_sham}. 
Let $V_{\rm d}$ and $V_{\rm s}$ represent the comoving volumes
of the observed galaxy sample and the simulations respectively. 
%Let us assume that the comoving volumes of the observed galaxy  sample and the simulation are given by $V_{\rm d}$ and $V_{\rm s}$ respectively.
Assuming that the covariance of the estimated 2PCF is inversely 
proportional to the volume (see \citealt{feldman94,tegmark97}), 
one can define an effective model covariance matrix in the following way:
\beq 
\widehat{C}_{\rm m} \simeq \widehat{C}_{\rm m, eff} = \frac{V_{\rm d}}{V_{\rm m}} \widehat{C}_{\rm d}
\eeq
Then one can write down an effective total covariance matrix as:
\beq 
\widehat{C} \simeq \Big(1+ \frac{V_{\rm d}}{V_{\rm m}}\Big)\widehat{C}_{\rm d},
\label{eq:volume}
\eeq
where the prefactor $\Big(1+ \frac{V_{\rm d}}{V_{\rm m}}\Big)$ accounts for the modeling uncertainties due to the finite simulation volume.

%\footnote{{\color{darkgreen} CHH: I don't think this footnote is necessary (see e-mail comment)} 
%Note that we have also considered computing the jackknife covariance matrix of the mock catalogs, but this procedure needs to be done every time a new set of HOD parameters is sampled in MCMC. This makes the MCMC analysis computationally prohibitive. Therefore, we use the approximation given by equation~\ref{eq:volume} to account for the model uncertainties.}.}

The advantage of using these measurements is that the effects of fiber collision systematic errors on the two-point statistics are corrected for (with the method described in \citealt{guo2012}), and therefore, these measurements provide accurate small scale clustering measurements. The assembly bias parameters introduced in section \ref{sec:method} can have a 10-percent level impacts on galaxy clustering (\citealt{decorated}). Presence of assembly bias in the satellite population impacts the very small-scale clustering (\citealt{decorated}). Moreover as \citet{sunayama2016} demonstrates, the scale-dependence of the halo assembly bias has a pronounced bump in the 1-halo to 2-halo transition regime (1$\sim$2 $h^{-1}\rm{Mpc}$). This scale can be impacted by fiber collision systematics. Precise investigation of the possible impact of this signal on the galaxy clustering modeling requires accurate measurements of 2PCF on small scales. 
The method of \citet{guo2012} is able to recover the true $\wpp$ with  $\sim 6\%$ accuracy in small scales ($r_{\rm{p}} = 0.1 \; h^{-1}\rm{Mpc}$) and with $\sim 2.5\%$ at relatively large scales $r_{\rm{p}} \sim 30 \; h^{-1}\rm{Mpc}$. 

Note that the comoving volume of the $N$-body simulation used in this investigation is 64 $\times 10^{6} \; h^{-3}\rm{Mpc}^{3}$ which is larger than the comoving volume of all the luminosity-threshold samples in the SDSS data considered in this study except the two most luminous samples. The comoving volumes of the $M_{\rm r}<-21, \; 21.5$ samples are 71.74 and 134.65 (in units of $10^{6} h^{-3}\rm{Mpc}^{3}$) respectively. Since we are not studying very large scale clustering ($r_{p,\; \rm max} \leq 25 \; h^{-1}\rm{Mpc}$), using a slightly smaller box for those samples is justified.

\section{Analysis}\label{sec:analysis}

\subsection{Inference setup}\label{subsec:analysis}

Given the SDSS measurements described in Section \ref{sec:data}, we aim to constrain the HOD model without assembly bias (described in \ref{subsubsec:hod}), and the HOD model with assembly bias (described in \ref{subsubsec:decorated}) for each luminosity-threshold sample, by sampling from the posterior probability distribution $p(\theta|d) \propto p(d|\theta) \pi(\theta)$ where $\theta$ denotes the parameter vector and $d$ denotes the data vector. In the standard HOD modeling $\theta$ is given by
\beq
\theta = \{ \mmin,\;\sigmam,\;\mzero,\; \alpha,\;\mone \},
\eeq
and in the HOD modeling with assembly bias we have 
\begin{eqnarray}
\theta = \{\mmin,\; \sigmam, \;\mzero, \;\alpha, \;\mone,\;\acen,\; \asat \},
\end{eqnarray}
Furthermore, data (denoted by $d$) is the combination of $[n_{g}, w_{p}(r_{p})]$. The negative log-likelihood (assuming negligible covariance between $n_{g}$ and $w_{p}(r_{p})$) is given by
\begin{eqnarray}
-2\ln p(d|\theta) &=& \frac{[n^{\rm data}_{g}-n^{\rm model}_{g}]^{2}}{\sigma_{n}^{2}} + \Delta \wpp^{\rm T}\widehat{C^{-1}}\Delta \wpp \; + \; \rm{const.},
\label{eq:lnlike_wp}
\end{eqnarray}
where $\Delta \wpp$ is a 12 dimensional vector, $\Delta \wpp(\rpp) = \wpp^{\rm data}(\rpp)-\wpp^{\rm model}(\rpp)$, and  $\widehat{C^{-1}}$ is the estimate of the inverse covariance matrix that is related to the inverse of the 
    total covariance matrix (Eq.~\ref{eq:volume})
 $\widehat{C}^{-1}$, following \citet{hartlap2007}:
\beq
\widehat{C^{-1}} = \frac{N -d - 2}{N -1} \; \widehat{C}^{-1},
\eeq
where $N=400$ is the number of the jackknife samples, and $d=12$ is the length of the data vector $w_{\rm p}$. Another important ingredient of our analysis is specification of the prior probabilities $\pi(\theta)$ over the parameters of the halo occupation models considered in this study. For both models, we use uniform flat priors for all the parameters. The prior ranges are specified in the Table \ref{tab:prior}. Note that a uniform prior between -1 and 1 is chosen for assembly bias parameters since these parameters are, by definition, bounded between -1 and 1. 
\begin{table}
\begin{center}
  \label{tab:prior}
  \caption{{\bf Prior Specifications}: The prior probability distribution 
  and its range for each of the parameters. 
  All mass parameters are in unit of $h^{-1}M_\odot$. The parameters marked by $*$ are only used in the Heaviside Assembly bias modeling and by definition are bounded between -1 and 1.}
\begin{tabular}{@{}lllll}
\\ \hline 
    Parameter & & Prior & & Range \\ \hline
  $\alpha$ & & Uniform & & [0.85, 1.45] \\
  $\sigmam$ & & Uniform & &  [0.05, 1.5] \\
  $\mzero$   & & Uniform & &  [10.0, 14.5] \\
  $\mmin$ & &   Uniform & &  [10.98, 14.0] \\
  $\mone$ & & Uniform & & [11.5, 15.0] \\ 
  $\acen^{*}$ & & Uniform & & [-1.0, 1.0] \\
  $\asat^{*}$ & & Uniform & & [-1.0, 1.0] \\
 \hline
  \end{tabular}
\end{center}
\end{table}

For sampling from the posterior probability, given the likelihood function (see equation \ref{eq:lnlike_wp}) and the prior probability distributions (see Table \ref{tab:prior}), we use the affine-invariant ensemble MCMC sampler (\citealt{goodmanweare}) and its implementation $\mathtt{emcee}$ (\citealt{emcee}). 
In particular, we run the $\mathtt{emcee}$ code with 
20 walkers and we run the chains for at least 12000 iterations. We discard the first one-third part of the chains as burn-in samples and use the reminder of the chains as production MCMC chains. Furthermore, we perform auto-correlation convergence test (\citealt{goodmanweare}) to ensure that the MCMC chains have reached convergence.

Once the posterior probability distributions are obtained, we estimate the corresponding minimum $\chi^{2}$ (or equivalently the maximum likelihood) of each HOD model for each luminosity-threshold sample. Finding the minimum $\chi^{2}$ is done using the $\mathtt{scipy}$ implementation of the $\mathtt{BFGS}$ algorithm \citep{bfgs}. The $\mathtt{BFGS}$ parameter bounds are given by the 68$\%$ confidence intervals of the posterior probability distributions. 

\section{Results and Discussion}\label{sec:result}

\subsection{Constraints and Interpretations}
\begin{figure*}[p]~\\
\begin{center}
\includegraphics[width=0.8\textwidth]{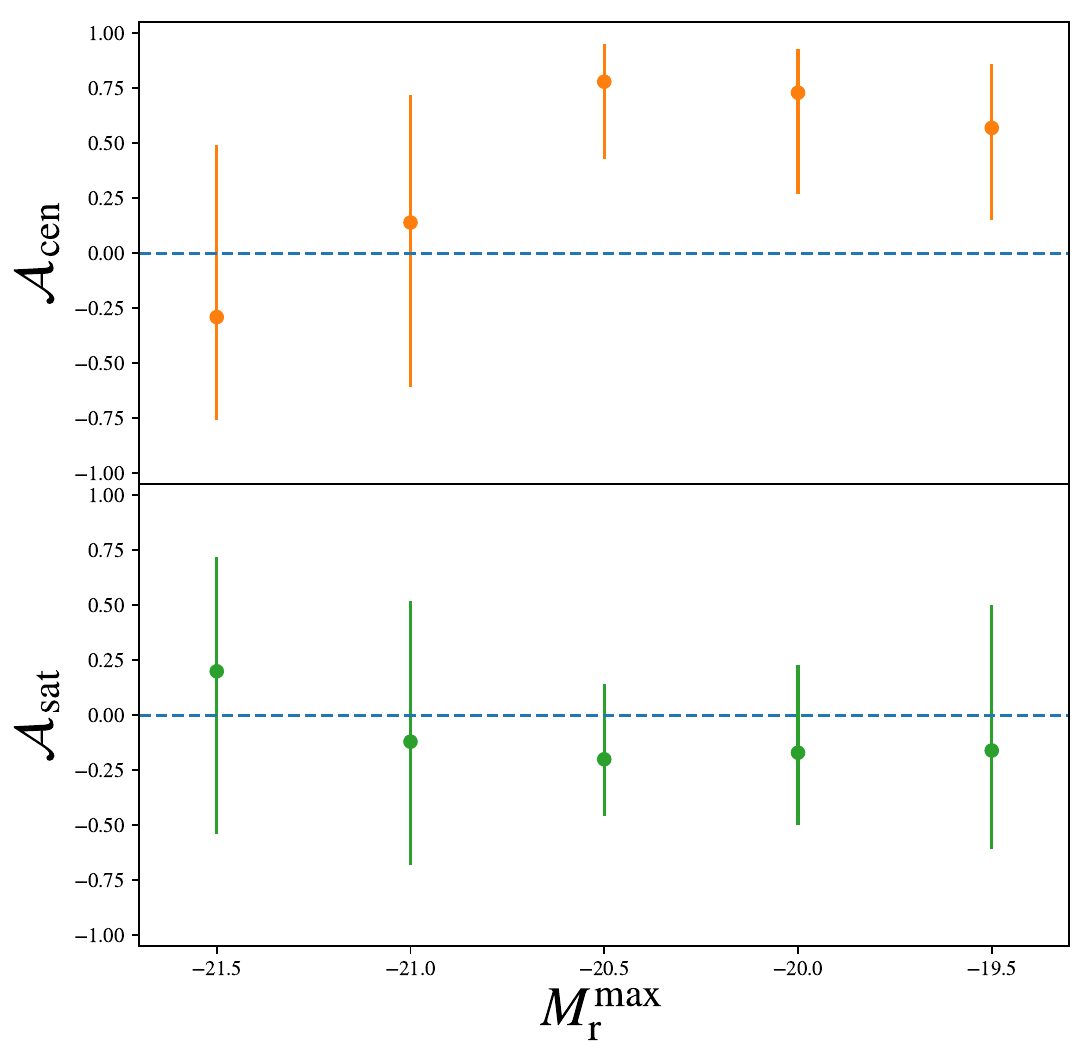}
\caption{Constraints on the central assembly bias $\acen$ (Top panel) and the satellite assembly bias $\asat$ (Bottom panel) parameters. The $\acen$ constraints for the $M_{\rm r} < -20.5, -20, -19.5$ samples favor positive values of $\acen$ with the tightest constraint coming from the $M_{\rm r} < -20.5$ sample. All the $\asat$ constraints are consistent with zero.}
\label{fig:bias}
\end{center}
\end{figure*}

%Our constraints on the assembly bias parameters fall into two main categories. First, the satellite 
%assembly bias parameter $\asat$ and the second, the central assembly bias parameter $\acen$. 
%%% Re-worded
    In this section, we present the constraints derived for the two assembly 
    bias parameters: the satellite assembly bias parameter ($\asat$) and the 
    central assembly bias parameter $\acen$. 
%%%
As shown in Figure \ref{fig:bias}, for all the five luminosity-threshold samples in the SDSS DR7 data, our constraints on the parameter $\asat$ are 
consistent with zero. On the other hand, our constraints on the parameter $\acen$--- albeit not tightly constrained--- show a trend which can be summarized as the following. In the most luminous galaxy samples, i.e. $M_{\rm r} < -21.5$ and $M_{\rm r} < -21$, $\acen$ is poorly constrained and the constraints are equivalent to zero. As we investigate less luminous samples, $M_{\rm r} < -20.5 , -20, -19.5$, our constraints on $\acen$ shift toward positive values, with the $M_{\rm r}<-20.5$ sample favoring the highest values for $\acen$. 

According to \citet{decorated}, the underlying theoretical considerations for explaining assembly bias (as formulated in the decorated HOD model) of the central and satellite galaxies are different. The large scale clustering---or the two halo term in the galaxy clustering---is mainly governed by the clustering of the central galaxies. The central galaxy clustering can be thought as the weighted average over the halo clustering. The large scale bias of the dark matter halo clustering depends not only on mass, but also on the other properties of halos beyond mass, such as concentration (\citealt{weschler2006,gao2007,miyatake2016}), spin (\citealt{gao2007}), formation time (\citealt{gao2007, li2008}), and maximum circular velocity of the halo $V_{\rm max}$ (\citealt{sunayama2016}).

In particular, findings of \citet{weschler2006} and \citet{sunayama2016}
have demonstrated that for halos with mass bellow the collapse mass ($M \leq M_{\rm col} \simeq 10^{12.5}\; M_{\odot} $) the large scale bias of high-$V_{\rm max}$ (or equivalently high-$c$ halos at a fixed halo mass) is larger than that of the low-$V_{\rm max}$ (low-$c$ halos). This signal reverses and weakens for the high mass halos ($M \geq M_{\rm col}$). Note that the halo concentration traces the maximum circular velocity $V_{\rm max}$ such that halos with higher $V_{max}$ have higher concentration and vice versa (see \citealt{prada2012}). For halos described by NFW profile, at a fixed halo mass, halos with higher values of concentration have higher values of $V_{\rm max}$.

Furthermore, investigation of the scale dependence of halo assembly bias has shown that the ratio of the bias of high-$V_{\rm max}$ halos and the low-$V_{\rm max}$ halos has a bump-like feature in the quasi-linear scales $\sim$ 0.5 $\mathrm{Mpc}h^{-1}-5\mathrm{Mpc}h^{-1}$. From the theoretical standpoint, this phenomenon has been attributed to a population of distinct halos with $M \sim 10^{11.7} \; h^{-1}M_{\odot}$ at the present time that are close to the most massive groups and clusters (\citealt{sunayama2016}, and see \citealt{more2016} for the observational investigation of this signal by means of galaxy-galaxy lensing). The clustering of these population of halos is therefore dictated by that of the massive halos. Note that this scale-dependent bias feature vanishes in high mass halos. 

Consequently, at a fixed halo mass less than $M_{\rm col}$, assignment of more central galaxies to the high-$c$ halos (higher expected number of central galaxies in the high-$c$ halos) gives rise to a boost in the galaxy clustering in the linear scales as well as in regimes corresponding to the one-halo to two-halo transition. For the more massive halos ($M \geq M_{\rm col}$), we expect the large-scale clustering boost to reverse sign, and the quasi-linear bump feature to vanish.  

\begin{figure*}[p]~\\
\begin{center}
\includegraphics[width=\textwidth]{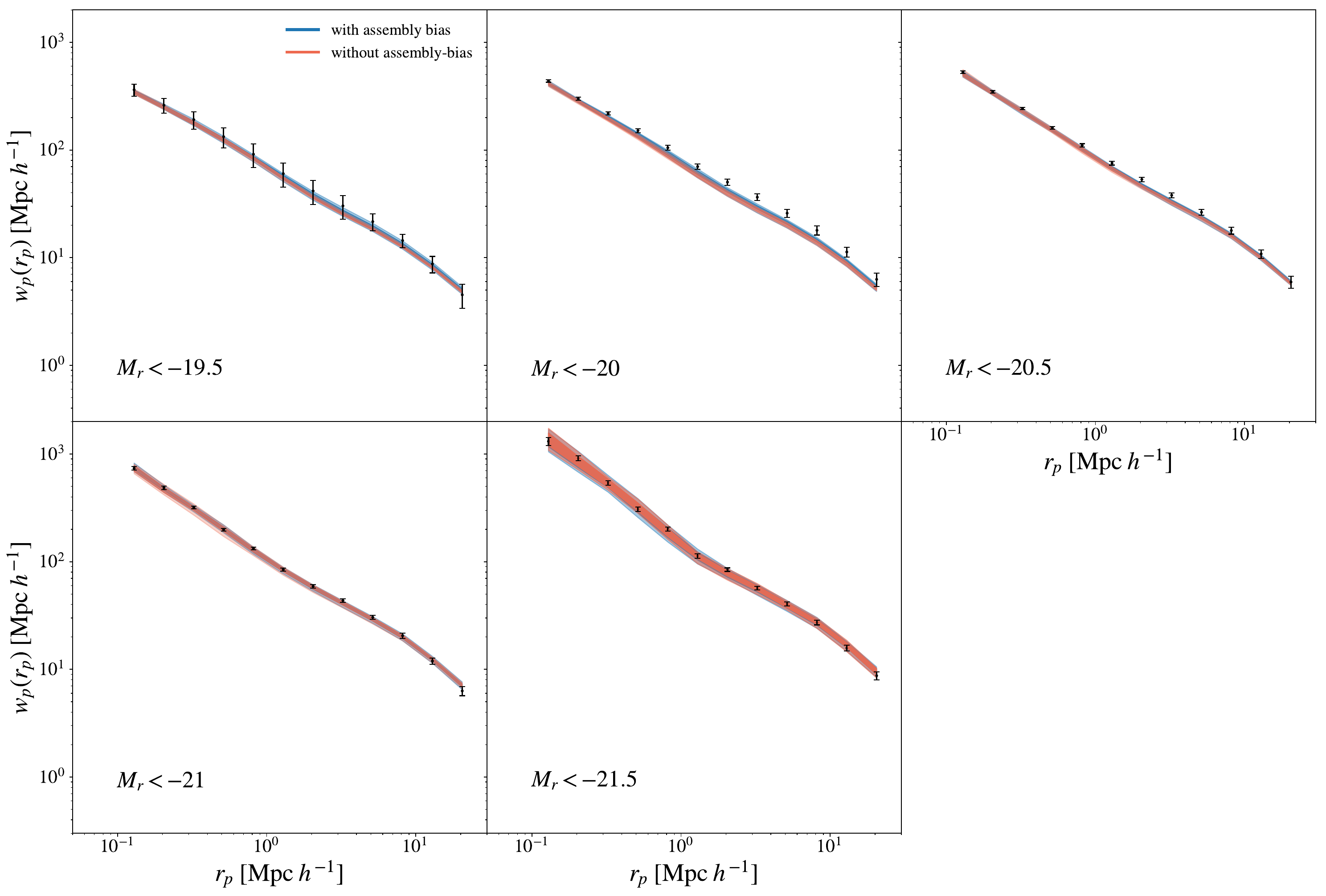}
\caption{Comparison between the posterior predictions of $w_{p}(r_{p})$ and the SDSS $w_{p}(r_{p})$ measurements. Predictions from the standard HOD model (HOD model with assembly bias) are shown in red (blue). The Dark and light shaded regions mark the 68$\%$ and the 95$\%$ confidence intervals. The errorbars are from the diagonal elements of the covariance matrix of the observations.}
\label{fig:wpmodel}
\end{center}
\end{figure*}

Figure \ref{fig:wpmodel} demonstrates the 68$\%$ and 95$\%$ posterior predictions for the projected 2PCF $w_{p}$ from the occupation model without assembly bias (shown in red) and the occupation model with assembly bias (shown in blue) for all five luminosity-threshold samples. For the brightest galaxies, $M_{r} < -21.5$ and $M_{r} < -21.0$, the posterior prediction of $w_{p}$ from the two models are consistent with one another. Note that these galaxies reside in the most massive halos ($M > M_{\rm col}$) for which the scale-dependence of the halo assembly bias and the difference between the large-scale bias of the high-$c$ and low-$c$ halos is negligible. 

\begin{figure*}[p]~\\
\begin{center}
\includegraphics[width=\textwidth]{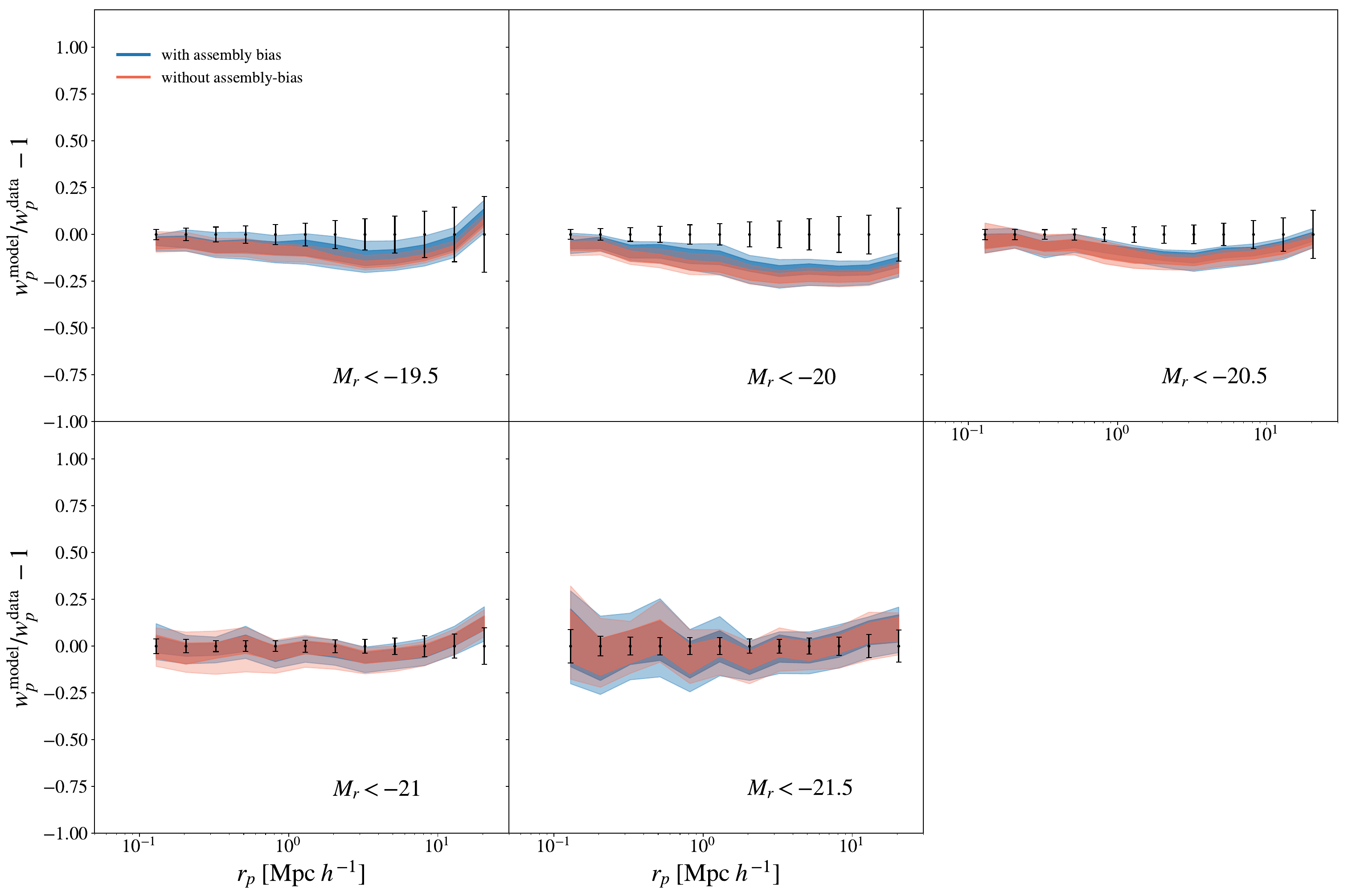}
\caption{Same as Figure \ref{fig:wpmodel}, but showing the fractional difference between the posterior predictions and the observed projected 2PCF for all the luminosity threshold samples. In all luminosity threshold samples, predictions of the two models for small scale clustering are consistent. In the samples that favor more positive values of the central assembly bias parameter ($M_{\rm r}<-19.5,-20,-20.5$), modeling of the intermediate and large scale clustering is slightly improved.}
\label{fig:wpres}
\end{center}
\end{figure*}

Figure \ref{fig:wpres} shows the fractional difference between the 68$\%$ and 95$\%$ posterior predictions of $w_p$ and the SDSS data. It is evident from Figure \ref{fig:wpres} that some improvement on modeling the clustering of the samples of $L_{\star}$ and slightly less brighter than $L_{\star}$ galaxies can be achieved by employing the more complex halo occupation model with assembly bias. As a result of apportioning more central galaxies to the high-$c$ halos relative to the low-$c$ halos, in the samples with luminosity thresholds of $M_{\rm r}<-20.5, -20, -19.5$, the posterior predictions for $w_{p}$ are slightly improved in the intermediate scales ($1\sim 2 \; \mathrm{Mpc} h^{-1}$) and large scales. This can be also noted in significantly lower $\chi^{2}$ values---at the cost more model flexibility and higher degrees of freedom---achieved by the assembly bias model in these luminosity-threshold samples (see Table \ref{tab:constraints}).  

%In the sample of galaxies with the luminosity threshold $M_{\rm r} < -19, -18.5$, the constraints on $\acen$ become consistent with zero with the tendency towards positive values for the $M_{\rm r}<-19$ sample and towards more negative values for the $M_{\rm r }<-18.5$ sample. As shown in Figures \ref{fig:wpmodel} and \ref{fig:wpres}, for the $M_{\rm r}<-19$ ($M_{\rm r}<-18.5$) sample this results in slightly higher (lower) posterior predictions for $w_{p}$ in the intermediate toward large scales. Overall, for these two samples, the assembly bias parameters remain largely unconstrained and the decorated HOD model does not yield better $\chi^{2}$ values. 

%Finally, In the faintest sample ($M_{\rm r} < -18$), negative constraints on the parameter $\acen$ results in higher expected number of centrals in the low-$c$ halos, which at a fixed halo mass, cluster less strongly. This affects both the large scale bias and the intermediate regimes as a result of the scale-dependent bump feature (see Figures \ref{fig:wpmodel} and \ref{fig:wpres}). Furthermore, the model with assembly bias provides better fit to the SDSS data in this luminosity-threshold sample. 

The luminosity dependent trend in the constraints on the central assembly bias for the six dimmest samples can be attributed to the fact that the halo concentration is highly correlated with the maximum circular velocity $V_{\rm max}$ which is a tracer of the potential well of dark matter halos (\citealt{prada2012}). In a dark matter halo described by an NFW profile, the depth of the gravitational potential well of dark matter halos can be directly measured by the maximum circular velocity $V_{\rm max}$ (\citealt{vmax_potential}). In particular, the magnitude of the potential well at the center of an NFW halo---where the central galaxy is assumed to reside---scales as $V_{\rm max}^{2}$. More specifically, we have:
\beq
\Phi(r=0) = -\Big(\frac{V_{\rm max}}{0.465}\Big)^{2},
\eeq
 where $\Phi(r=0)$ is the central potential of an NFW profile. Note that $V_{\rm max}$ is also the quantity often used in the abundance matching technique in which the luminosity of galaxies is monotonically matched to $V_{\rm max}$ (see for example \citealt{reddick2013,lehman2015, hod_vs_sham, halodemographic}). The trend between the constraint on $\acen$ and the luminosity threshold of the samples may suggest that at a \emph{fixed} \emph{halo} \emph{mass}, the central galaxies in the samples ($M_{\rm max}<$-19.5, -20, -20.5),  have a tendency to reside in dark matter halos with deeper gravitational potential well. 

The satellite assembly bias can only significantly alter the galaxy clustering at small-to-intermediate scales. Assigning more satellite galaxies to lower (or higher) concentration halos affects the one-halo term through increasing the satellite-satellite pair counts $\langle N_{\rm s} N_{\rm s} \rangle$ (\citealt{decorated}). This results in boosting the small-scale clustering. But as pointed out by \citet{decorated}, the amount by which small-scale clustering increases also depends on the sign of the central assembly bias parameter $\acen$. Formation history of the halos can lead to the dependence of the abundance of subhalos on halo concentration (\citealt{zentner2005, mao2015}) at fixed halo mass, and since the occupation of the satellite galaxies is related to the abundance of subhalos, the satellite occupation may depend on halo concentration. 

However, our results suggest that for all the luminosity-threshold samples considered in this study, the satellite assembly bias parameter is largely unconstrained and consistent with zero. We do not expect the galaxy clustering data to be a sufficient statistics for obtaining constraints on the satellite assembly bias. Group statistics probes the high mass end of the galaxy--halo connection and is sensitive to the parameters governing the satellite population (see \citealt{sham_gmf, 2016arXiv160701782H}). Therefore these measurements may shed some light on potential presence of assembly bias in satellite population.   

\begin{figure*}[p]~\\
\begin{center}
\includegraphics[width=0.9\textwidth]{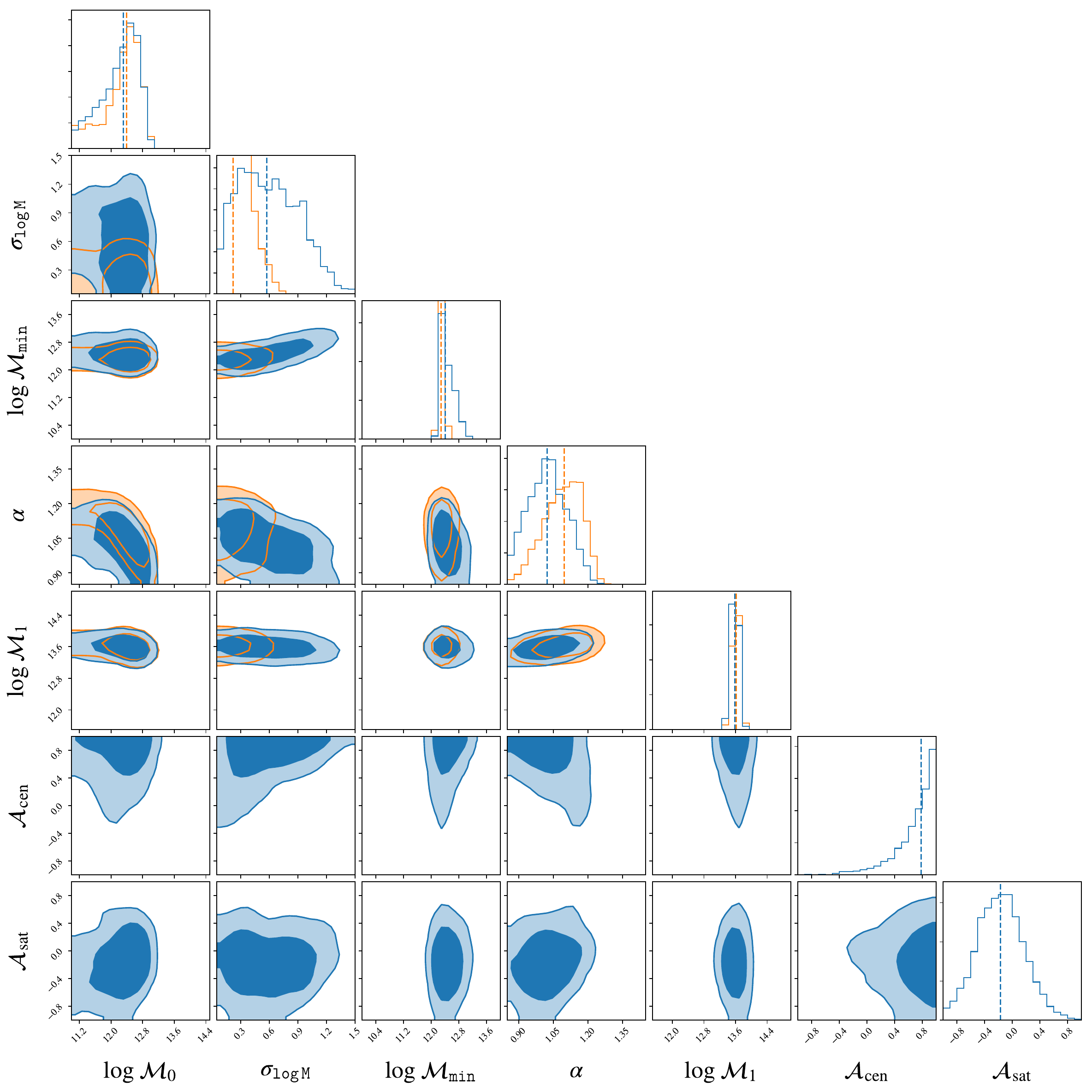}
\caption{An example of posterior probability distribution over the parameters of the standard HOD model with no assembly bias (shown with yellow), and the HOD model with assembly bias (shown in blue). These constraints are obtained from the clustering measurements of the $M_{\rm r} < -20.5$ luminosity threshold sample. The dark (light) blue shaded regions show the 68$\%$ (95 $\%$) confidence intervals. The constraints on $\acen$ and $\asat$ show positive correlation between the central occupation and the halo concentration at fixed halo mass, and lack of correlation between the satellite occupation and halo concentration at fixed halo mass.}
\label{fig:posterior}
\end{center}
\end{figure*}

It is important to note that the halo mass range in which the central assembly bias $\acen$ affects the central galaxy population is the mass range in which the condition $0<\langle N_{\rm c}|M \rangle<1$ is met. Consequently, larger scatter parameter $\sigmam$ increases the dynamical mass range in which assembly bias affects the galaxy clustering. Note that in the luminosity regimes for which we obtain tighter constraints on $\acen$, the best-estimate values of the scatter parameter $\sigmam$ appear to be higher in the model with assembly bias. This is evident in Figure \ref{fig:posterior}. The model with assembly bias tends to push $\sigmam$ to higher values. This can be attributed to the tendency of this model to increase the effective dynamical mass range of central assembly bias (\citealt{decorated}).

As shown in Table \ref{tab:constraints}, in the HOD model with assembly bias, the constraints found on the scatter parameter are not tight. This is in keeping with the results of \citet{guo2015} which uses the same SDSS measurements and finds that the scatter parameter remains largely unconstrained when only $n_{g}$ and $w_{p}$ are used as observables. Note that scatter is better constrained for the most luminous galaxy samples (this is attributed to the steep dependence of the halo bias and halo mass function on halo mass in the high mass end). But since these samples live in the most massive halos, we do not expect the tighter constraints on scatter to help us constrain the central assembly bias parameter. \citet{guo2015} shows that by employing additional measurements such as the monopole ($\xi_{0}$), quadruple ($\xi_{2}$), and hexadecapole ($\xi_{4}$), one can obtain tighter constraints on the scatter parameter. Tightening the constraints on the scatter parameter can lead to more precise inference of the central assembly bias parameter.    

As shown in Table \ref{tab:constraints}, our constraints on the underlying standard HOD model obtained from the model with assembly bias and the model without assembly bias are in good agreement. The only cases in which there are mild tensions between the constraints found from the two models on the underlying HOD parameters, are the $M_{\rm r}<-20$ and the $M_{\rm r}<-20.5$ samples. However, these tensions are still within one-sigma level. For instance, Figure \ref{fig:posterior} shows that in the $M_{\rm r}<-20.5$ sample, the constraint on the parameter $\alpha$ found from the model without assembly bias favors slightly higher values than the constraint found from the model with assembly bias. Also the scatter parameter $\sigmam$ is more tightly constrained in the standard HOD model. However, it is important to emphasize that these constraints are still in agreement with each other within a one-sigma level. 

\citet{arz2014} shows that in the mock catalogs that exhibit significant levels of assembly bias, using a simple mass-only occupation model can lead to considerable biases in inference of the galaxy--halo connection parameters. Although we cannot rule out moderate levels of assembly bias in our findings, we do not find any considerable discrepancy between the two models in terms of estimating the underlying HOD parameters. 
 
A few galaxy--halo connection methods have been proposed in the literature that give rise to assembly bias in the galaxy population. 
%%%% CHH: (Ref. Report Minor #5) re-worded and Conroy et al. (2006) included in citations. 
    \citet{arz2014} demonstrates that the abundance matching techniques based on $V_{\rm max }$ 
    \citep{conroy:2006aa,hw2013,reddick2013} exhibit some levels of assembly bias. 
We aim to provide a comparison between the impact of assembly bias on clustering in these mock catalogs and the mock catalogs predicted from our constraints on the decorated HOD model for $L_{\star}$-type galaxies. In particular, we consider the abundance matching catalogs produced by \citet{hw2013}. These catalogs have been extensively studied for examining potential systematic effects of galaxy assembly bias on cosmological (\citealt{edHOD-weinberg}) and halo occupation (\citealt{arz2014}) parameter inferences. 

This abundance matching catalog was built based on the $\mathtt{Bolshoi}$ $N$-body simulation (\citealt{Klypin2011}) using the adaptive refinement tree code (ART \citealt{art}). The Box size for this simulation is 250 $h^{-1} \rm{Mpc}$, the number of simulation particles is 2048$^3$, the mass per simulation particle $m_{\rm p}$ is $1.35 \times 10^{8} \; h^{-1} M_{\odot}$, and the gravitational softening length $\epsilon$ is 1 $h^{-1} \rm{kpc}$. The halos and subhalos in this simulation are identified using the ROCKSTAR algorithm (\citealt{rockstar}). 
The \citealt{hw2013} catalogs make use of $V_{\rm peak}$ (maximum $V_{\rm max}$ throughout the assembly history of halo) as the subhalo property to be matched to galaxy luminosity. 

As noted by \citet{arz2014} and \citet{edHOD-weinberg}, these galaxy mock catalogs show significant levels of assembly bias in the central galaxy population. This has been demonstrated by investigating the difference in $w_{p}$ between the randomized mock catalogs and the original mock catalogs. Randomization is performed in a procedure described in \citet{arz2014} which we briefly summarize here: First, halos are divided into different bins of halo mass with width of 0.1 dex. Then all central galaxies are shuffled among all halos within each bin. Once the centrals have been shuffled, within each bin, the satellite systems are shuffled among all halos in that mass bin, preserving their relative distance to the center of halo. This procedure preserves the HOD, but erases any dependence of the galaxy population on the assembly history of halos. Therefore, assembly bias is erased in the randomized galaxy catalog. 

\begin{figure*}[p]~\\
\begin{center}
\includegraphics[width=\textwidth]{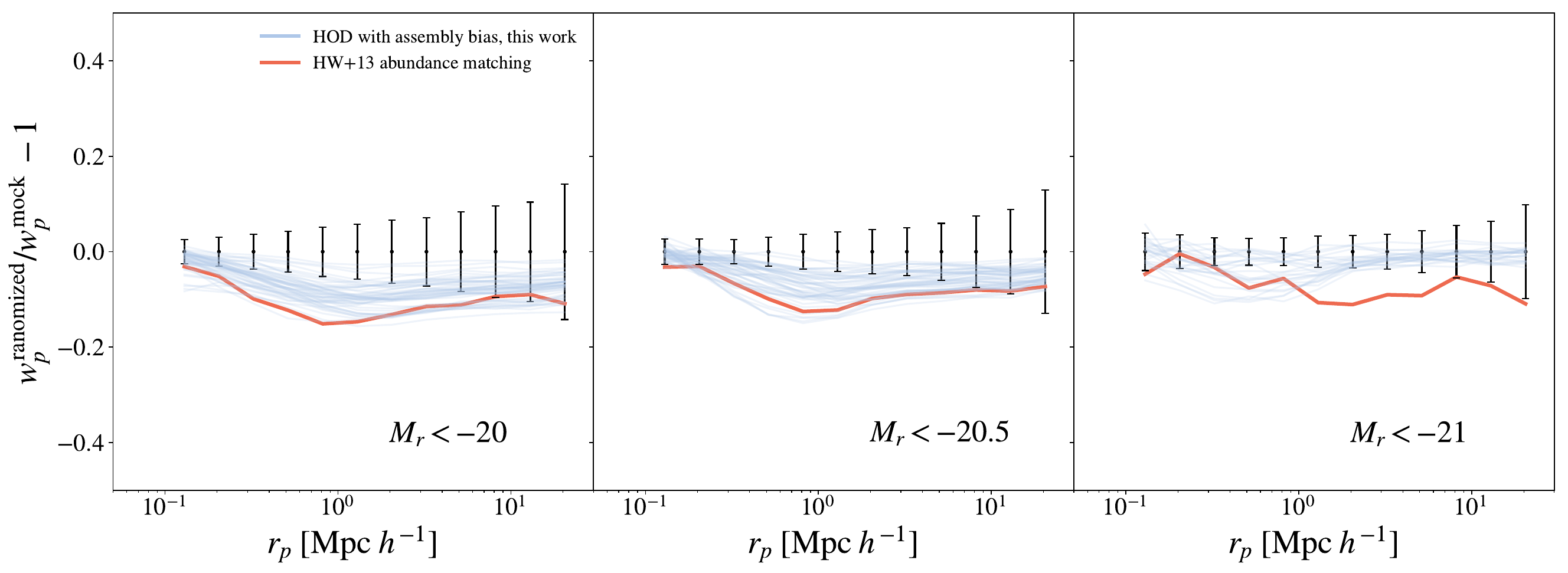}
 \caption{Demonstration of the relative difference in $w_{p}$ between randomized and non-randomized catalogs for different luminosity threshold samples: $M_{\mathrm r}<-20,-20.5,-21$. The errorbars are from the diagonal elements of the covariance matrix. The blue lines correspond to the random draws from the $68\%$ confidence intervals of the posterior probability (summarized in Table \ref{tab:constraints}) over the parameters of the HOD model with assembly bias. The red line corresponds to the subhalo abundance matching catalog (\citealt{hw2013,hearin2014}). Our constraints favor \emph{more} \emph{moderate} levels of the impact of assembly bias on galaxy clustering than the levels seen in the abundance matching mock catalogs. Within both models, the small scale clustering remains unaltered after randomizing the catalogs, signaling the lack of correlation between the satellite occupation and the halo concentration at a fixed mass in the two models.}
\label{fig:randomized}
\end{center}
\end{figure*}

For $L_{\star}$ galaxies, the difference in $w_{p}$ between the randomized and the original catalogs of \citet{hw2013} is shown in Figure \ref{fig:randomized} with the red curves. As demonstrated in Figure \ref{fig:randomized} (and as previously noted by \citealt{arz2014,edHOD-weinberg}), the relative difference in $w_{p}$ is only significant in relatively large scales ($r_{p} > 1 \mathrm{Mpc} \; h^{-1}$). This implies that in these catalogs, only the central occupation is affected by assembly bias. This is in agreement with our findings. 

Furthermore, we investigate whether the impact of assembly bias on galaxy clustering predicted by our findings are in agreement with the abundance matching catalogs of \citet{hw2013}. First, we make random draws from the 68$\%$ confidence intervals of the posterior probability distribution function over the parameters of the model with assembly bias. Then, we create mock catalogs with these random draws, and then we compute the difference in $w_{p}$ between the randomized catalogs and the original catalogs. The relative difference in $w_{p}$ predicted from our constraints are shown with blue curves in Figure \ref{fig:randomized}. 

We note that our findings follow the same trend. That is, we see negligible difference in the small scale clustering and more considerable differences in $w_{p}$ on larger scales ($r_{p} > 1\; \mathrm{Mpc}h^{-1}$). Similar to findings of \citet{arz2014}, \citet{lehman2015}, and \citet{edHOD-weinberg}, we see that the impact of assembly bias on galaxy clustering becomes less significant in brighter galaxy samples. Furthermore, we notice that our mock catalogs favor more moderate changes in galaxy clustering as a result of assembly bias. 

\begin{table*}
\begin{center}
  \label{tab:constraints}
 \caption{{\bf Constraints}: Constraints on the parameters of the HOD models with and without assembly bias. 
All mass parameters are in unit of $h^{-1}M_\odot$. The best-estimates and the error bars correspond to the 50$\%$ quantile and 68$\%$ confidence intervals obtained from the marginalized posterior probability pdfs. The last column is $\chi^{2}$ per degrees of freedom ($\mathrm{dof}$), where $\mathrm{dof} = N_{\rm data} - N_{\rm par}$.}
\begin{tabular}{@{}lllllllllllllllllllllll}
\hline 
   $M_{\rm r,lim}$ & $\mmin$ & $\sigmam$ & $\mzero$ & $\alpha$ &  $\mone$ & $\acen$ & $\asat$ & $\chi^{2}/\rm{dof}$ \\  \hline
  \\ 
 %-18 & $11.56^{+0.21}_{-0.25}$ &  $1.05^{+0.31}_{-0.52}$ & $10.84^{+0.77}_{-0.59}$ & $0.98^{+0.05}_{-0.05}$ &  $12.50^{+0.10}_{-0.10}$ & $-$ & $-$ & 14.51/8\\\\
     
  %-18 & $11.53^{+0.23}_{-0.21}$ &  $1.08^{+0.28}_{-0.51}$ & $10.86^{+0.81}_{-0.62}$ & $0.97^{+0.05}_{-0.04}$ &  $12.56^{+0.09}_{-0.10}$ & $-0.67^{+0.55}_{-0.25}$ & $-0.30^{+1.09}_{-0.54}$ & 7.52/6\\ \\
     
%-18.5 & $11.67^{+0.29}_{-0.25}$ &  $0.83^{+0.45}_{-0.53}$ & $10.85^{+0.64}_{-0.60}$ & $1.02^{+0.04}_{-0.04}$ &  $12.69^{+0.08}_{-0.08}$ & $-$ & $-$ & 6.17/8\\ \\
    
%-18.5 & $11.60^{+0.31}_{-0.20}$ &  $0.74^{+0.49}_{-0.46}$ & $10.73^{+0.69}_{-0.51}$ & $1.01^{+0.05}_{-0.05}$ &  $12.72^{+0.09}_{-0.10}$ & $0.02^{+0.67}_{-0.62}$ & $0.07^{+0.53}_{-0.59}$ & 6.23/6\\ \\

%-19 & $11.74^{+0.37}_{-0.18}$ &  $0.62^{+0.52}_{-0.39}$ & $10.82^{+0.62}_{-0.56}$ & $1.04^{+0.04}_{-0.04}$ &  $12.87^{+0.09}_{-0.09}$ & $-$ & $-$ & 8.69/8\\ \\

%-19 & $11.71^{+0.37}_{-0.16}$ &  $0.58^{+0.53}_{-0.38}$ & $10.75^{+0.66}_{-0.52}$ & $1.03^{+0.04}_{-0.05}$ &  $12.90^{+0.10}_{-0.09}$ & $0.36^{+0.44}_{-0.62}$ & $-0.01^{+0.56}_{-0.54}$ & 8.87/6\\ \\

%%%%%%%wp%%%%%

-19.5 & $11.82^{+0.24}_{-0.16}$ &  $0.66^{+0.53}_{-0.43}$ & $11.51^{+0.40}_{-0.36}$ & $1.06^{+0.03}_{-0.04}$ &  $13.03^{+0.07}_{-0.07}$ & $-$ & $-$ & 6.62/8 \\ \\

-19.5 & $11.88^{+0.21}_{-0.19}$ &  $0.81^{+0.44}_{-0.49}$ & $11.64^{+0.40}_{-0.44}$ & $1.01^{+0.05}_{-0.06}$ &  $13.02^{+0.10}_{-0.09}$ & $0.57^{+0.29}_{-0.42}$ & $-0.16^{+0.66}_{-0.45}$ & 5.98/6\\ \\

-20 & $12.03^{+0.23}_{-0.10}$ &  $0.41^{+0.42}_{-0.25}$ & $11.82^{+0.47}_{-0.54}$ & $1.07^{+0.04}_{-0.06}$ &  $13.3^{+0.09}_{-0.09}$ & $-$ & $-$ & 20.68/8\\ \\

-20 & $12.21^{+0.31}_{-0.24}$ &  $0.79^{+0.47}_{-0.51}$ & $11.77^{+0.53}_{-0.53}$ & $1.00^{+0.05}_{-0.06}$ &  $13.26^{+0.09}_{-0.09}$ & $0.73^{+0.20}_{-0.46}$ & $-0.17^{+0.40}_{-0.33}$ & 17.28/6\\ \\

-20.5 & $12.29^{+0.07}_{-0.05}$ &  $0.23^{+0.17}_{-0.12}$ & $12.37^{+0.32}_{-0.68}$ & $1.1^{+0.07}_{-0.1}$ &  $13.62^{+0.08}_{-0.09}$ & $-$ & $-$ & 10.02/8\\ \\

-20.5 & $12.36^{+0.16}_{-0.09}$ &  $0.48^{+0.25}_{-0.29}$ & $12.45^{+0.28}_{-0.53}$ & $1.03^{+0.08}_{-0.09}$ &  $13.59^{+0.09}_{-0.08}$ & $0.78^{+0.17}_{-0.35}$ & $-0.20^{+0.34}_{-0.26}$ & 4.7/6\\ \\

-21 & $12.72^{+0.11}_{-0.07}$ &  $0.31^{+0.21}_{-0.18}$ & $12.59^{+0.5}_{-0.95}$ & $1.19^{+0.09}_{-0.17}$ &  $14.01^{+0.08}_{-0.11}$ & $-$ & $-$ & 6.22/8\\ \\

-21 & $12.72^{+0.1}_{-0.07}$ &  $0.31^{+0.21}_{-0.17}$ & $12.47^{+0.54}_{-0.92}$ & $1.17^{+0.09}_{-0.15}$ &  $14.05^{+0.07}_{-0.1}$ & $0.14^{+0.58}_{-0.75}$ & $-0.12^{+0.64}_{-0.56}$ & 5.9/6\\ \\

-21.5 & $13.39^{+0.14}_{-0.11}$ &  $0.56^{+0.15}_{-0.17}$ & $12.87^{+0.65}_{-1.23}$ & $1.26^{+0.13}_{-0.24}$ &  $14.51^{+0.07}_{-0.09}$ & $-$ & $-$ & 2.7/8\\ \\

-21.5 & $13.42^{+0.13}_{-0.13}$ &  $0.58^{+0.16}_{-0.18}$ & $12.59^{+0.78}_{-1.07}$ & $1.25^{+0.14}_{-0.24}$ &  $14.55^{+0.08}_{-0.08}$ & $-0.29^{+0.78}_{-0.47}$ & $0.2^{+0.52}_{-0.74}$ & 3.24/6\\ \\
                
 \hline
  \end{tabular}
\end{center}
\end{table*}

\subsection{model comparison}
%%% CHH: reworded. 
Using the constraints on the model with assembly bias and without, 
we want to address the question of whether assembly bias is supported
by galaxy clustering observations. %We want to address this question that whether the constraints on the model with assembly bias and the model without assembly bias given the galaxy clustering data lead us to claim that assembly bias is strongly supported by the observations or not. 
Within the standard HOD framework, the distribution of the galaxies is modeled using a simple description based on the $mass$-only ansatz: $P(N|M)$. The decorated HOD model however, provides a more complex description of the data by adding a secondary halo property (halo concentration in this study) and a more flexible occupation model: $P(N|M,c)$. 

In order to investigate whether observations demand higher level of 
model complexity, we compare the models with and without assembly bias.

In particular, we make use of two simple methods for model comparison: \emph{Akaike Information Criterion} (AIC, \citealt{aic} , see \citealt{gelmanic} for detailed discussion on AIC), and \emph{Bayesian Information Criteria} (BIC, \citealt{bic}). BIC and AIC are more computationally tractable than alternative approaches such as computing the fully marginalized likelihood. The underlying assumption of these information criteria is that models that yield higher likelihoods are more preferable, but at the same time, models with more flexibility are penalized.

Suppose $\mathcal{L}^{\star}$ is the maximum likelihood achieved by
the model, $N_{\rm par}$ is the number of free parameters in the 
model, and $N_{\rm data}$ is the number of data points in the data 
set. Then we have
\begin{eqnarray}
\rm{BIC}&=& -2\;\ln \mathcal{L}^{\star} + N_{\rm par}\;\ln N_{\rm data} \; , \label{eq:bic} \\
\rm{AIC}&=& -2\;\ln \mathcal{L}^{\star} + 2N_{\rm par} \; . \label{eq:aic}
%         &+& 2\;\frac{N_{\rm par}(N_{\rm par}+1)}{N_{\rm data}-N_{\rm par}-1} \; . \label{eq:aic}
\end{eqnarray}
Given a data set, models with lower value of AIC and BIC are more desired. 
That is, higher model complexity (larger $N_{\rm par}$) is favored 
only if $\mathcal{L}^{\star}$ is sufficiently higher.

%in order for the higher model complexity (given by $N_{\rm par}$) to be justified, $\mathcal{L}^{\star}$ must be sufficiently higher. Therefore in this formulation, models that deliver lower information criteria scores are more preferable. 

\begin{figure*}[p]~\\
\begin{center}
\includegraphics[width=0.8\textwidth]{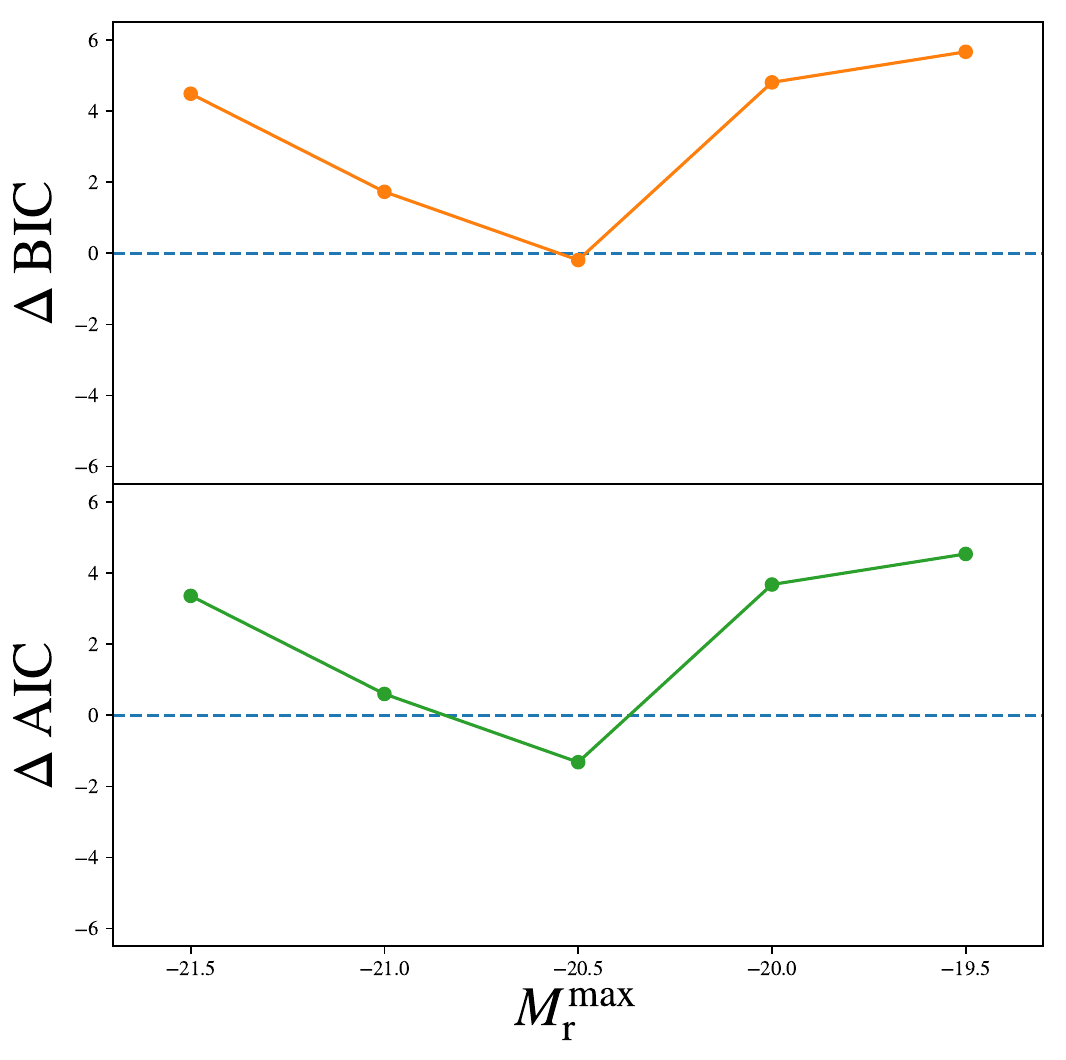}
\caption{Difference in the information criteria between the HOD model with assembly bias and the model without assembly bias. $\bf{Top}$: $\Delta$BIC = BIC(with assembly bias) - BIC(without assembly bias). $\bf{Bottom}$: $\Delta$AIC = AIC(with assembly bias) - AIC(without assembly bias). According to BIC (AIC), the more complex model with assembly bias is favored once $\Delta$BIC$<0$ ($\Delta$AIC$<0$). Both $\Delta$BIC and $\Delta$AIC are lower for the samples with tighter constraints over the central assembly bias parameter $\acen$, with $\Delta$BIC and $\Delta$AIC being (marginally) negative only for the $M_{\rm r}<-20.5$ sample that yield the strongest constraint on $\acen$.} 
\label{fig:ic}
\end{center}
\end{figure*}

Figure \ref{fig:ic} shows the comparison between the BIC and AIC 
scores for the model with assembly bias and the model without 
assembly bias.
The more complex decorated HOD model is favored by the information criteria if $\Delta$IC$<0$, and strongly favored if $\Delta$IC$<-5$. With the exception of the $M_{\rm r}<-20.5$ sample,
the model without assembly bias is still preferred by both information criteria. Although including
assembly bias improves the fit to the observed clustering, overall these 
improvements are not enough to justify adding the extra assembly bias 
parameters to the model.

In the $M_{\rm r}<-20.5$ sample where both AIC and BIC scores improve 
for the assembly bias model, we have the greatest improvement in terms
of the minimum $\chi^2$. This is also the sample with the tightest 
constraints on the parameters $\acen$ and $\asat$. While the
information criteria favor the model with assembly bias in this 
sample, both $\Delta$AIC and $\Delta$BIC are only marginally negative. 
Therefore, even in this sample, the improvement in the achieved
$\mathcal{L}^{\star}$ by additional parameters is not enough to make 
the decorated model strongly favored by AIC and BIC.

%This supports our intuition that AIC and BIC penalize unconstrained parameters. Also note that, the difference between both BIC and AIC scores are marginal.

\section{Summary and Conclusion}\label{sec:summary}

In this investigation, we provide constraints on the concentration-dependence of halo occupation for a wide range of galaxy luminosities in the SDSS data. In particular, the modeling is done in the context of the decorated HOD model \citet{decorated}, and the data used in this investigation is the projected 2PCF measurements published by \citet{guo2015}. We make use of the $\mathtt{SMDP}$ high resolution $N$-body simulation with a volume and resolution suitable for studying the clustering of galaxies in the SDSS DR7 main galaxy sample.

%that enables us to reliably perform inference for the faint galaxy samples in the SDSS DR7 catalog that live in low mass halos, and the brightest galaxy samples that occupy large comoving volumes. 

Our findings suggest that the satellite assembly bias remains consistent with zero. However, our constraints on the central assembly bias parameter exhibit a trend with the luminosity limits of the galaxy samples. For the brightest samples, central assembly bias is consistent with zero, which is in agreement with this picture that the halo assembly bias becomes negligible for the most massive halos. 

For the $M_{\rm r}<-20.5, -20, -19.5$ samples, at a fixed halo mass, we find positive correlation between the central population and halo concentration at fixed halo mass. 
%For $M_{\rm r}<-19, -18.5$ we find no correlation, and for the faintest sample, we find negative correlation. 
Given the large scale halo assembly bias and the scale-dependent feature of assembly bias in the quasi-linear scales, our constraints on the more flexible HOD model lead to improvement in modeling the galaxy clustering. However, we do not find these improvements to be sufficient to lower the information criteria scores associated with the more complex model. 
%The exceptions are the $M_{\rm r}<-20, -18$ luminosity-threshold samples for which we find the strongest constraints on the central assembly bias. For these two samples, the HOD model with assembly bias yields lower BIC score than the model without assembly bias.
The exceptions is the $M_{\rm r}<-20.5$ luminosity-threshold sample for which we find the strongest constraints on the central assembly bias. For this sample, the HOD model with assembly bias yields lower BIC score than the model without assembly bias.

We compare the impact of assembly bias on galaxy clustering between the catalogs constructed from our results and the abundance matching catalogs presented in \citet{hw2013,arz2014}. We demonstrate that the effect of assembly bias on galaxy clustering predicted from our results is similar to (but more moderate than) the effects seen in the abundance matching catalogs of \citet{hw2013}. That is, assembly bias mostly affects the large scales and the quasi-linear clustering and the small scale clustering remains unaltered. In addition, the effect of assembly bias on galaxy clustering vanishes in the brightest galaxy samples.

Moreover, we repeat our inference using the $\mathtt{BolshoiP}$ simulation. 
We find that our findings based on the $\mathtt{BolshoiP}$ simulation are consistent with %constraints reported by \citet{zentner2016} (in the $M_{\rm r}<-21, -20.5, -20, -19.5, -19$ 
constraints reported by \citet{zentner2016} (in the $M_{\rm r}<-21, -20.5, -20, -19.5$
luminosity-threshold samples) that predicts positive satellite assembly bias (correlation between the expected number of satellites and $V_{\rm max}$ at fixed host halo mass) for the %$M_{\rm r}<-19, -20.5$ samples. However, we note that the results based on the $\mathtt{SMDP}$ 
$M_{\rm r}<-20.5$ samples. However, we note that the results based on the $\mathtt{SMDP}$
simulation are more consistent with the picture provided by the previous models based on the abundance matching technique (e.g. \citealt{arz2014,lehman2015}). That is, only the large-scale clustering, governed by the centrals, is affected by assembly bias.

\section*{Acknowledgments}

We are grateful to David~W.~Hogg, Alex~I.~Malz, Andrew~Hearin, Andrew~Zentner, Chia-Hsun~Chuang, Michael~R.~Blanton, and Kilian~Walsh for discussions related to this work. This work was supported by the NSF grant AST-1517237. All the computations in this work were carried out in the New York University High Performance Computing Mercer facility. We thank Shenglong~Wang, the administrator of the NYU HPC center for his continuous and consistent support throughout the completion of this study. 

The CosmoSim database used in this paper is a service by the Leibniz-Institute for Astrophysics Potsdam (AIP). The MultiDark database was developed in cooperation with the Spanish MultiDark Consolider Project CSD2009-00064. The authors gratefully acknowledge the Gauss Centre for Supercomputing e.V. (www.gauss-centre.eu) and the Partnership for Advanced Supercomputing in Europe (PRACE, www.prace-ri.eu) for funding the MultiDark simulation project by providing computing time on the GCS Supercomputer SuperMUC at Leibniz Supercomputing Centre (LRZ, www.lrz.de).

All of the code written for this project is available in an open-source
code repository at \url{https://github.com/mjvakili/gambly}. The SDSS clustering measurements and the covariance matrices used in this work are available at \url{http://sdss4.shao.ac.cn/guoh/files/wpxi_measurements_Guo15.tar.gz}. Description of the $\mathtt{SMDP}$ and the $\mathtt{BolshoiP}$ halo catalogs used in this investigation can be found at \url{https://www.cosmosim.org/cms/simulations}. The RockStar halo catalogs of the $\mathtt{SMDP}$ and the $\mathtt{BolshoiP}$ simulations are publicly available at \url{http://yun.ucsc.edu/sims/SMDPL/hlists/index.html} and \url{http://yun.ucsc.edu/sims/Bolshoi_Planck/hlists/index.html} respectively. We thank Peter Behroozi for making the halo catalogs publicly available. In this work we have made use of the publicly available codes: corner (\citealt{corner}), emcee (\citealt{emcee}), halotools (\citealt{halotools}), Corrfunc (\citealt{corrfunc}), and changtools (\url{https://github.com/changhoonhahn/ChangTools}). The abundance matching mock catalogs used in this study are available at (\url{http://logrus.uchicago.edu/~aphearin/}).

\clearpage

%%%%%%%%%%%%%%%% FIGURES %%%%%%%%%%%%%%

%%%%%%%%%%%%%%%%%%%%%%%%%%%%%%%%%%%%%%%%%%%%%%%%%%%%%%%%
% ACEN and ASAT constraints
%%%%%%%%%%%%%%%%%%%%%%%%%%%%%%%%%%%%%%%%%%%%%%%%%%%%%%%%

%\clearpage

%%%%%%%%%%%%%%%%%%%%%%%%%%%%%%%%%%%%%%%%%%%%%%%%%%%%%%%%
% WP MODEL
%%%%%%%%%%%%%%%%%%%%%%%%%%%%%%%%%%%%%%%%%%%%%%%%%%%%%%%%
%\clearpage
%%%%%%%%%%%%%%%%%%%%%%%%%%%%%%%%%%%%%%%%%%%%%%%%%%%%%%%%
% WPRES
%%%%%%%%%%%%%%%%%%%%%%%%%%%%%%%%%%%%%%%%%%%%%%%%%%%%%%%%

%\clearpage

%%%%%%%%%%%%%%%%%%%%%%%%%%%%%%%%%%%%%%%%%%%%%%%%%%%%%%%%
% WP RANDOMIZED
%%%%%%%%%%%%%%%%%%%%%%%%%%%%%%%%%%%%%%%%%%%%%%%%%%%%%%%%

%\clearpage

%%%%%%%%%%%%%%%%%%%%%%%%%%%%%%%%%%%%%%%%%%%%%%%%%%%%%%%%
% Information Criteria
%%%%%%%%%%%%%%%%%%%%%%%%%%%%%%%%%%%%%%%%%%%%%%%%%%%%%%%%

%\clearpage

%%%%%%%%%%%%%%%%%%%%%%%%%%%%%%%%%%%%%%%%%%%%%%%%%%%%%%%%
% BIAS COMPARISON
%%%%%%%%%%%%%%%%%%%%%%%%%%%%%%%%%%%%%%%%%%%%%%%%%%%%%%%%

%\clearpage

%%%%%%%%%%%%%%%%%%%%%%%%%%%%%%%%%%%%%%%%%%%%%%%%%%%%%%%%
% ASAT DISCREPANCY
%%%%%%%%%%%%%%%%%%%%%%%%%%%%%%%%%%%%%%%%%%%%%%%%%%%%%%%%
%\begin{figure*}[p]~\\
%\begin{center}
%\includegraphics[width=\textwidth]{hist_comparison.pdf}
%\caption{Constraints over the satellite assembly bias parameters from luminosity-threshold samples $M_{\rm r}<-19 ,\; -20.5$, for two different simulations: $\mathtt{BolshoiP}$ (yellow), and $\mathtt{SMDPL}$ (green). The $\asat$ constraints found using the $\mathtt{BolshoiP}$ simulation favor more positive values of $\asat$, while the constraints found using the $\mathtt{SMDP}$ simulation favor zero satellite assembly bias.}
%\label{fig:asat_comparison}
%\end{center}
%\end{figure*}

%\clearpage

%%%%%%%%%%%%%%%%%%%%%%%%%%%%%%%%%%%%%%%%%%%%%%%%%%%%%%%%
% EXAMPLE POSTERIOR PDF
%%%%%%%%%%%%%%%%%%%%%%%%%%%%%%%%%%%%%%%%%%%%%%%%%%%%%%%%

\bibliographystyle{yahapj}
\bibliography{assemble}
\appendix
\subsection{choice of simulation}

Given the SDSS clustering measurements described in Section \ref{sec:data}, We repeat the inference of the assembly bias parameters $\asat$ and $\acen$ with the $\mathtt{BolshoiP}$ simulation (\citealt{smallmultidark}). This $N$-body simulation is carried out with similar setting as the $\mathtt{Bolshoi}$ simulation with the exception that in the $\mathtt{BolshoiP}$ simulation, Planck cosmology is adapted and the mass per simulation particle is $1.49\times 10^8 \; h^{-1} M_{\odot}$. 

\begin{figure*}[p]~\\
\begin{center}
\includegraphics[width=0.8\textwidth]{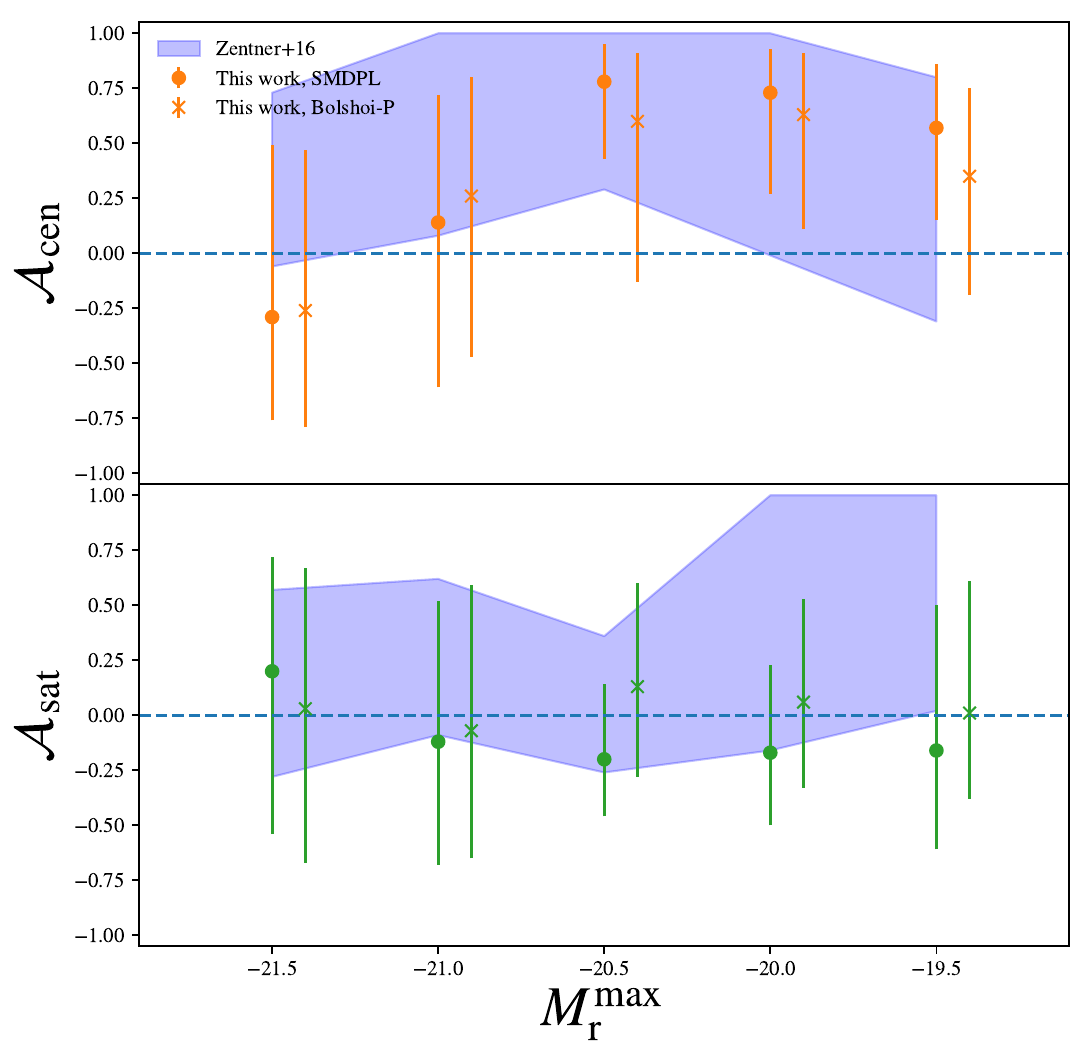}
\caption{Comparison between the constraints on the assembly bias parameters $\acen$ (shown in the top panel) and $\asat$ (shown in the bottom panel) for different simulations: $\mathtt{SMDP}$ (shown with circle), and $\mathtt{BolshoiP}$ (shown with cross). The errorbars mark the 68$\%$ uncertainty over the parameters. Shaded blue regions show the upper and lower bounds reported by \citet{zentner2016} that uses the $\mathtt{BolshoiP}$ and clustering measurements of \citet{zehavi2011}. For the confidence intervals corresponding to the shaded blue regions, we refer the readers to Table 2 of \citet{zentner2016}. The central assembly bias constraints found from the two simulations are consistent, with the constraints for from the $\mathtt{SMDP}$ simulation being tighter for the most luminous samples. The constraints on $\asat$ from the two simulations are largely in agreement.}
\label{fig:bias_comparison}
\end{center}
\end{figure*}

The summary of constraints are shown in Figure \ref{fig:bias_comparison}. In Figure \ref{fig:bias_comparison}, the constraints from the $\mathtt{SMDP}$ and the $\mathtt{BolshoiP}$ simulations are shown with circles and crosses respectively. Additionally, the upper and lower bounds on the inferred parameters reported by \citet{zentner2016} are shown in shaded blue regions. 
In the case of central assembly bias, our constraints based on the $\mathtt{SMDP}$ and the $\mathtt{BolshoiP}$ simulations are consistent. For the luminosity-thresholds samples $M_{\rm r}<-20.5, -20, -19.5$, where the central assembly bias parameters are strongly positive, the constraints obtained from the $\mathtt{SMDP}$ simulation are tighter. For the three faintest luminosity threshold samples, our constraints are fully consistent with those of \citet{zentner2016}. For the brightest samples ($M_{\rm r}<-21.5, -21$) however, the \citet{zentner2016} constraints on $\acen$ are slightly higher than ours.
%$In the case of central assembly bias, all three constraints are consistent. For the luminosity-thresholds samples $M_{\rm r}<-20.5, -20, -19.5$, where the central assembly bias parameters are strongly positive, the constraints obtained from the $\mathtt{SMDP}$ simulation are tighter. 
  
%In the case of the satellite assembly bias parameters however, constraints from the $\mathtt{BolshoiP}$ simulation for the luminosity threshold samples $M_{\rm r}<-20.5, -19$ favor more positive values of the parameter, while our constraints from the $\mathtt{SMDP}$ simulation for these two luminosity thresholds favor zero satellite assembly bias. As it is shown in the lower panel of Figure \ref{fig:bias_comparison}, our constraints from the $\mathtt{BolshoiP}$ simulations for the $M_{\rm r}<-21, -20.5, -20, -19.5, -19$ samples are consistent with the lower and upper bounds (shown with the shaded blue region) reported by \citet{zentner2016} that uses the same simulation but different $w_{p}$ measurements (\citealt{zehavi2011}). Therefore, there is some discrepancy between our $\asat$ constraints using the $\mathtt{SMDP}$ simulation and the $\mathtt{BolshoiP}$ simulations. 
In the case of satellite assembly bias, our constraints based on the $\mathtt{SMDP}$ and the $\mathtt{BolshoiP}$ simulations are fully consistent with each other and consistent with $\asat = 0$. For the luminosity-thresholds samples $M_{\rm r}<-20.5, -20$, where the estimated central assembly bias parameters are strongly positive, the $\asat$ constraints obtained from the $\mathtt{SMDP}$ simulation are tighter. For all the luminosity-threshold samples except for the brightest sample, the \citet{zentner2016} constraints on $\asat$ are slightly higher than ours. But given the large uncertainties on $\asat$, our constraints on this parameter seem to be consistent with those of \citet{zentner2016}.
%For the $M_{\rm r}<-19, -20.5$ samples, the marginalized posterior PDFs over $\asat$ from the two simulations are shown in Figure \ref{fig:asat_comparison}. Note that $\asat$ is poorly constrained in both simulations and for both luminosity-threshold samples. For the $M_{\rm r}<-19$ sample, considering how poorly constrained the parameters are, the discrepancy between the constraints is not very stark. Note that the tension is still at a one-sigma level. The discrepancy however, becomes more pronounced in the $M_{\rm r}<-20.5$ sample. 

In terms of the effect of assembly bias on galaxy clustering, note that mocks created using the inferred parameters with the $\mathtt{SMDP}$ simulation show the same behavior as we observe in the abundance matching catalogs presented in \citet{arz2014} and \citet{lehman2015}. That is, the difference in $w_{p}$ between the mock catalogs and the randomized catalogs is mostly on large scales where the clustering is governed by the central galaxies. That is, the impact of assembly bias on the satellite occupation is negligible and only the central occupation is affected. 

\end{document}